\documentclass[journal]{IEEEtran}
\usepackage{amsmath}
\usepackage{amssymb}
\usepackage{amsfonts}
\usepackage{amsthm}
\usepackage{eucal}
\usepackage{array}
\usepackage{bm}
\usepackage{graphicx}
\usepackage{fancyhdr}
\usepackage{stfloats}
\usepackage{multirow}
\usepackage{multicol}
\usepackage{color}
\usepackage{algorithm}
\usepackage{algorithmic}
\usepackage{cite}
\usepackage[bookmarks=false]{}
\usepackage{mathbbol}
\usepackage{latexsym}
\usepackage{cite}
\usepackage{url}
\usepackage{stfloats}
\usepackage{mathrsfs}
\usepackage{setspace}
\usepackage{booktabs}
\usepackage{titlesec}
\usepackage{lipsum}
\usepackage[top = 0.7in, bottom=0.95in,left=0.65in,right=0.65in]{geometry}
\makeatletter
\renewcommand{\maketag@@@}[1]{\hbox{\m@th\normalsize\normalfont#1}}%
\makeatother
\theoremstyle{plain}

\ifCLASSINFOpdf
\else
\fi
\usepackage[small]{subfigure}
\usepackage{caption}
\begin{document}
\clearpage
\pagestyle{empty}
\title{Joint LED Selection and Precoding Optimization for Multiple-User Multiple-Cell VLC Systems}
\author{\IEEEauthorblockN{Yang Yang, \emph{Member, IEEE}, Yujie Yang, \emph{Student Member, IEEE},
Mingzhe Chen, \emph{Member, IEEE}, Chunyan Feng,   \emph{Senior Member, IEEE}, Hailun Xia, \emph{Member, IEEE}, Shuguang Cui, \emph{Fellow, IEEE} and H. Vincent Poor, \emph{Life Fellow, IEEE}}
\vspace{-0.3cm}
 \thanks{Y. Yang, Y. Yang, C. Feng and H. Xia are with the Beijing Key Laboratory of Network System Architecture and Convergence, School of Information and Communication Engineering, Beijing University of Posts and Telecommunications, Beijing 100876, China (e-mail: yangyang01@bupt.edu.cn; yangyujie@bupt.edu.cn; cyfeng@bupt.edu.cn; xiahailun@bupt.edu.cn).}
 \thanks{M. Chen and H. V. Poor are with the Department of Electrical and Computer Engineering, Princeton University, Princeton, NJ, 08544, USA (e-mail: mingzhec@princeton.edu and poor@princeton.edu).}
 \thanks{S. Cui is currently with the Shenzhen Research Institute of Big Data and Future Network of Intelligence Institute (FNii), the Chinese University of Hong Kong, Shenzhen, China, 518172 (e-mail: shuguangcui@cuhk.edu.cn).}
\thanks{ This work was supported by National Natural Science Foundation of China (61871047), National Natural Science Foundation of China (61901047), and Beijing Natural Science Foundation (4204106). The work was supported in part by the National Key R{\&}D Program of China with grant No. 2018YFB1800800, by the Key Area R{\&}D Program of Guangdong Province with grant No. 2018B030338001, by Shenzhen Outstanding Talents Training Fund 202002, and by Guangdong Research Projects No. 2017ZT07X152 and No. 2019CX01X104.   
}
\thanks{Copyright (c) 2021 IEEE. Personal use of this material is permitted. However, permission to use this material for any other purposes must be obtained from the IEEE by sending a request to pubs-permissions@ieee.org.}
	
}
%
%
%
\maketitle
\thispagestyle{empty}
\begin{abstract}
 This paper proposes a hybrid dimming scheme based on joint LED selection and precoding design (TASP-HD) for multiple-user (MU) multiple-cell (MC) visible light communications (VLC) systems. In TASP-HD, both the LED selection and the precoding of each cell can be dynamically adjusted to reduce the intra- and inter-cell interferences while satisfying illumination constraints. First, a MU-MC-VLC system model is established, and then a sum-rate maximization problem under dimming level and illumination uniformity constraints is formulated. In this studied problem, the indices of activated LEDs and precoding matrices are optimized, which result in a complex non-convex mixed integer problem. To solve this problem, the original problem is separated into two subproblems. The first subproblem, which maximizes the sum-rate of users via optimizing the LED selection with a given precoding matrix, is a mixed integer problem solved by the penalty method. With the optimized LED selection matrix, the second subproblem which focuses on the maximization of the sum-rate via optimizing the precoding matrix is solved by the Lagrangian dual method. Finally, these two subproblems are iteratively solved to obtain a convergent solution. Simulation results verify that in a typical indoor scenario under a dimming level of 70\%, the mean bandwidth efficiency of TASP-HD is 4.8 bit/s/Hz and 7.13 bit/s/Hz greater than AD and DD, respectively.
\end{abstract}

\vspace{-0.2cm}
\section{Introduction}
\label{sec:intro}
\IEEEPARstart{V}{isible} light communication (VLC) with its abundant license-free spectrum has become a promising technology for high speed data transmission and accurate positioning \cite{MUsurvey,6G,yzh1}. Different from the radio frequency based communications \cite{Chen2019,MISO}, VLC uses light-emitting diodes (LEDs) as transmitters. Therefore, using VLC to service users, one must jointly consider both communications and illumination. The main purpose of dimming control is to enhance the communication performance of a VLC system. Dimming control reduces energy consumption and provides ecological benefits, and satisfies the users' subjective requirements such as mood adjustment. Therefore, dimming capability is an essential function of popular commercial off-the-shelf LEDs \cite{dimLED}. However, dimming capability can significantly affect the waveform of VLC signals, which further affects communication performance of a VLC system. Therefore, it is interest to design novel dimming control schemes that have advanced communication performance while being compatible with dimming capability.

There is considerable prior art on dimming control \cite{dd1,dd2,ad2,ad3,spadim1,hybdim1,hybdim3}, which can be classified into three categories: digital dimming (DD)\cite{dd1,dd2}, analog dimming (AD)\cite{ad2,ad3} and spatial dimming (SD)\cite{spadim1}. DD achieves dimming control by adjusting the duty cycle of the transmitted signals. However, using DD, the data rate of each user significantly depends on the duty cycle. Hence, the data rate will be restricted due to a small duty cycle. AD is simple and cost effective, while the amplitude of the signal is determined by the target dimming level, and thus the communication performance is limited by the dimming level due to the constraint of the limited dynamic range of LEDs \cite{ad3}. SD is proposed to achieve dimming control by adjusting the number of glared LEDs without altering the signal forms. However the dimming range and precision of SD are influenced by the number of available LEDs. Besides, several hybrid dimming (HD) schemes that incorporate two of the aforementioned dimming control schemes have been proposed in \cite{hybdim1,hybdim3}. HD has the advantages of reducing chromaticity shift and clipping noise compared with AD, and achieving precise dimming control without constraint on the number of LEDs when compared with SD. Though interesting, most of the existing dimming control schemes are designed for single cell scenarios \cite{dd1,dd2,ad2,ad3,spadim1,hybdim1,hybdim3}. In practice, multi-cell (MC) scenarios are more practical for indoor VLC scenarios such as office building.

Although the dimming schemes designed for single cell scenarios can be directly applied to each cell of MC scenarios, there is a paucity of studies on certain vital aspects of dimming schemes in MC scenarios. In particular, on the one hand, from the perspective of communication performance, independent design of each cell can result in both intra-cell interference (intra-CI) and inter-cell inference (inter-CI). Even though the intra-CI can be eliminated by precoding design, the inter-CI can be severed with per-cell precoding design. On the other hand, from the perspective of illumination, independent dimming control at each cell fails to consider the overall illumination uniformity, which is crucial to user comfort according to International Organization for Standardization (ISO) standard \cite{ISOuir}. In fact, the above mentioned two perspectives are closely related. For instance, an area covered by multiple LEDs could have significant inter-CI but the illumination uniformity is improved due to the constructive addition of multiple visible light. Therefore, it is necessary to study the dimming control schemes for MC scenarios. Fortunately, intra-CI is closely related to the precoding design while inter-CI is dependent on the selected activated LEDs. In addition, both precoding and the selection of activated LEDs affect the illumination of the system. Therefore, it is desirable to design a dimming control scheme based on the joint LED selection and precoding design.

The main contribution of this paper is an efficient framework that a multiple-user (MU) MC VLC system services the users with dimming support. The objective is to maximize the sum rate under certain dimming constraints via LED selection and precoding design. To the best of our knowledge, this is the first hybrid dimming control scheme for MU-MC-VLC systems$\footnote{The conference version of this paper has been accepted by 2020 IEEE Global Communications Conference.}$. The key contributions are listed as follows:

\noindent
\setlength{\hangindent}{1em}
$\bullet$ We construct an MU-MC multiple-input single-output (MISO) VLC system model, where multiple LEDs transmit signals to users, and each user is equipped with a photodiode (PD). In the MU-MC scenario, the overlap of multiple LEDs could deteriorate the intra-CI and inter-CI while enhancing the illumination uniformity. To investigate the trade-off between communications and illumination of the system, we formulate a joint LED selection and precoding matrix design problem whose goal is to maximize the sum-rate while satisfying the uniform illumination requirement. Since the elements of the LED selection matrix are all binary integers, this problem is a non-convex, mixed integer problem, which is nondeterministic polynomial time hard (NP-hard). Therefore, we develop an efficient suboptimal iterative algorithm that divides the problem into two subproblems.

\noindent
\setlength{\hangindent}{1em}
$\bullet$ We first propose a LED selection algorithm to solve a sum-rate maximization problem given the precoding matrix and the uniform illumination constraint. Since the elements of LED selection matrix are binary integers, this subproblem is a mixed integer problem. To solve this problem, the integer variables are slackened into continuous variables. Then we construct a penalty function to represent the optimal optimization problem, and show that the slackened problem has the same solutions as that of the original mixed integer problem. Finally, the algorithm is solved iteratively by an interior point method.

\noindent
\setlength{\hangindent}{1em}
$\bullet$ With a given LED selection, we then propose a precoding design scheme to solve the second subproblem that optimizes the sum-rate of users under the signal amplitude constraint. We first analyze the equation of the amplitude constraint and transform it into a convex function. Then, the second subproblem becomes a convex problem, which is solved by Lagrangian dual method.

The simulation results show that TASP-HD can achieve better performance than the conventional AD and DD schemes in terms of the illumination uniformity and mean bandwidth efficiency. In particular, in a typical indoor scenario under dimming level of 70\%, the mean bandwidth efficiency of TASP-HD is 4.8 bit/s/Hz and 7.13 bit/s/Hz greater than AD and DD, respectively.

The remainder of this paper is organized as follows. In Section II, the system model and the optimization problem are presented. Section III introduces the user-centric cell formation and Section IV illustrates the proposed TASP-HD. Section V provides numerical and simulation results on the performance of TASP-HD and makes comparasion with conventional dimming schemes. Finally, Section VI concludes this paper.

\emph{Notations:} Bold upper case letters represent matrices and blackboard bold letters represent sets. ${\boldsymbol{A}}^{\rm{T}}$ is the transpose of matrix ${\boldsymbol{A}}$, ${{\boldsymbol{A}}_{(i,j)}}$ is the element at the \textit{i}th row and \textit{j}th column, ${{\boldsymbol{A}}_{(k,:)}}$ is the \textit{k}th row vector of ${\boldsymbol{A}}$ and ${{\boldsymbol{A}}_{(k,:)}}$ is the \textit{k}th column of ${\boldsymbol{A}}$. ${\left\|  \cdot  \right\|_{\rm{1}}}$ is the $L_1$ norm operator, $\mathcal{R}$ is the real number sets, $E\left[  \cdot  \right]$ is expectation operator, $\left\lfloor \cdot \right\rfloor $ and $\left|  \cdot  \right|$ are round down operator and absolute value operator, respectively.

\begin{table}[t!]
		\vspace{0.3cm}
	\small
	\caption{List of Variables.}
	\begin{center}
		\begin{tabular*}{\hsize}{cc}
			\hline
			\textbf{Variable}& \textbf{Definition}\\
			\hline
			${N_{\rm{T}}}$   & Total number of LEDs \\			
			${N_{\rm{R}}}$   & Total number of users\\
			${N_{c,\rm{T}}}$ & Number of LEDs in the $c$th cell\\
			${N_{c,\rm{R}}}$ & Number of users in the $c$th cell\\		
			${n_t}$   & Total number of activated LEDs \\
			${{\boldsymbol{H}}_c}$   & Channel matrix in the $c$th cell\\
			${{\boldsymbol{H}}_{c,(i,:)}}$ & The channel between LEDs and the $\textit{i}$th user \\&in the $c$th cell \\
			${\boldsymbol{H}}_c^\dag$ & The generalized inverse matrix of ${\boldsymbol{H}}_c$\\
			${{\boldsymbol{W}}_c}$   & Precoding matrix in the $c$th cell\\	
			${{\boldsymbol{A}}_c}$   & LED selection matrix in the $c$th cell\\
			${{\boldsymbol{d}}_c}$   & Normalized PAM data vector of the $c$th cell\\
			${I_{\rm{B}}}$	 & Direct current bias added to transmit signal\\
			${x_{c,j}}$      & The transmitted signal from the $j$th LED to users \\&in the $c$th cell\\
			${y_{c,i}}$      & The received signal of the $i$th user in the $c$th cell\\
			$n_{c,i}$        & The additive white Gaussian noise (AWGN) of the $\textit{i}$th  \\&user in the $c$th cell	\\
			$\sigma _{c,i}^2$& Variance of AWGN  \\   		    	
			\hline
		\end{tabular*}
	\end{center}\label{tab1}

\end{table}
\vspace{0.4cm}
\section{System Model and Problem Formulation}
\vspace{0.cm}
\subsection{VLC Channel Model}
\begin{figure}[t!]
	\centerline{\includegraphics[width=1\linewidth]{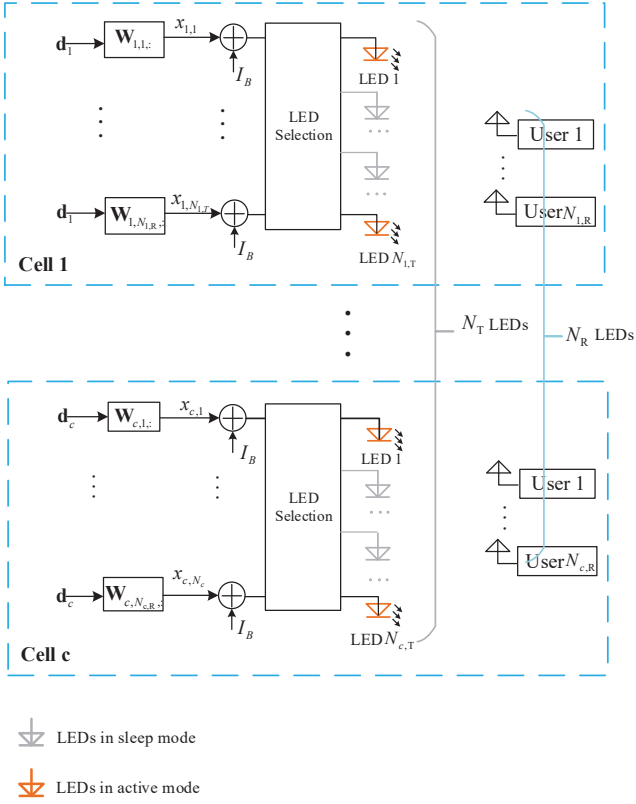}}
	\vspace{-0.cm}
	\caption{System model of TASP-HD in MU-MC-VLC system.}
	\label{block1}
\vspace{-0cm}
\end{figure}

Consider a MC-MU-MISO system model that consists of ${N_{\rm{T}}}$ LEDs, ${N_{\rm{R}}}$ users and ${N_c}$ cells is as shown in Fig. \ref{block1}. Each cell $c$ has ${N_{c,\rm{T}}}$ LEDs and ${N_{c,\rm{R}}}$ users with
\vspace{-0.mm}$\sum\limits_{c = 1}^{{N_c}}{{N_{c,\rm{T}}}}  = {N_{\rm{T}}}$, and $\sum\limits_{c = 1}^{{N_c}} {{N_{c,\rm{R}}}}  = {N_{\rm{R}}}$. Each user is equipped with a PD. For ease of reading, the notations of key system parameters are summarized in \textbf{Table I}. We assume that each LED obeys Lambertian beam distribution, and the channel between the \textit{i}th user and the \textit{j}th LED in the \textit{c}th cell is thus given by

\vspace{-0.cm}
\begin{footnotesize}
\begin{equation}
\hspace{-2mm}
\begin{array}{l}
{h_{c,i,j}} = \\
\left\{ {\begin{array}{*{20}{c}}
{\frac{{{A_r}\left( {l + 1} \right)}}{{2\pi d_{c,i,j}^2}}{{\cos }^l}({\phi _{c,i,j}}){T_s}({\psi _{c,i,j}})g({\psi _{c,i,j}})\cos ({\psi _{c,i,j}}),0 \le {\psi _{c,i,j}} \le \Psi, }\\
{0,\quad\quad\quad\quad\quad\quad\quad\quad\quad{\psi _{c,i,j}} > \Psi, }
\end{array}} \right.
\end{array} \label{eq2}
\end{equation}
\end{footnotesize}where ${A_r}$ is the detect area of the PD, $l =  - \frac{{\log (2)}}{{\log ({\Phi _{{1 \mathord{\left/
 {\vphantom {1 2}} \right.
 \kern-\nulldelimiterspace} 2}}})}}$ is the order of Lambertian emission determined by the semi-angle for half illuminance of the LED ${\Phi _{{1 \mathord{\left/
 {\vphantom {1 2}} \right.
 \kern-\nulldelimiterspace} 2}}}$. ${d_{c,i,j}}$, ${\phi _{c,i,j}}$, and ${\psi _{c,i,j}}$ are the distance, irradiance angle, and incidence angle between the $i$th user and the $j$th LED of the $c$th cell, respectively. ${\Psi}$ is the optical field-of-view (FOV) of the PD, ${{T_s}({\psi _{c,i,j}})}$ is the gain of optical filter and $g({\psi _{c,i,j}})$ is the gain of the optical concentrator defined in \cite{ad3}.
\newcounter{TempEqCnt}
\setcounter{TempEqCnt}{\value{equation}}
\setcounter{equation}{4}
\begin{figure*}
	\begin{align}
		{y_{c,i}} &= \gamma \varsigma \left[ {\begin{array}{*{20}{c}}
				{{{\boldsymbol{H}}_{1,(i,:)}},}&{{{\boldsymbol{H}}_{2,(i,:)}},}& \cdots &{{{\boldsymbol{H}}_{{N_c},(i,:)}}}
		\end{array}} \right]\left[ {\begin{array}{*{20}{c}}
				{{{\boldsymbol{x}}_1}}\\
				{{{\boldsymbol{x}}_2}}\\
				\vdots \\
				{{{\boldsymbol{x}}_{{N_c}}}}
		\end{array}} \right] + {n_{c,i}}\notag\\
		&= \gamma \varsigma \left( {{{\boldsymbol{H}}_{c,(i,:)}}{{\boldsymbol{W}}_{c,(:,i)}}{d_{c,i}} + {{\boldsymbol{H}}_{c,(i,:)}}\sum\limits_{k \in {{\mathcal{U}}_c},k \ne i}^{{N_{c,{\rm{R}}}}} {{{\boldsymbol{W}}_{c,(:,k)}}{d_{c,k}}}  + \sum\limits_{c' \ne c}^{{N_c}} {\sum\limits_{j \in {{\mathcal{U}}_{c'}}}^{{N_{c',{\rm{R}}}}} {{{\boldsymbol{H}}_{c',(i,:)}}{{\boldsymbol{W}}_{c',(:,j)}}{d_{c',j}}}  + {{\boldsymbol{H}}_c}{\boldsymbol{I}}_{\rm{B}}^c + {{\boldsymbol{H}}_{c'}}{\boldsymbol{I}}_{\rm{B}}^{c'}} } \right) + {n_{c,i}}\notag\\
		&= \gamma \varsigma \left( {{{\boldsymbol{H}}_{c,(i,:)}}{{\boldsymbol{W}}_{c,(:,i)}}{d_{c,i}} + \sum\limits_{c' \ne c}^{{N_c}} {\sum\limits_{j \in {{\mathcal{U}}_{c'}}}^{{N_{c',{\rm{R}}}}} {{{\boldsymbol{H}}_{c',(i,:)}}{{\boldsymbol{W}}_{c',(:,j)}}{d_{c',j}}} } } \right) + {n_{c,i}},\label{eq6}
	\end{align}
\vspace{-0.cm}
\end{figure*}
\setcounter{equation}{\value{TempEqCnt}}At the transmitter, the signals transmitted to the users within cell $c$ are precoded by ZF precoder given by ${{\boldsymbol{W}}_{c}}\in {\mathcal{R}^{{N_{c,\rm{T}}} \times {N_{c,\rm{R}}}}}$. Note that ZF can only eliminate intra-CI, but inter-CI still exists. To ensure that the amplitude of transmitted signal is within the dynamic range of LEDs, a direct current (DC) bias ${I_{\rm{B}}}$ is added. The transmit signal from the \textit{j}th LED to the users in the \textit{c}th cell is written as
\vspace{-0.cm}
\begin{equation}
{x_{c,j}} = {{\boldsymbol{W}}_{c,(j,:)}}{{\boldsymbol{d}}_c} + {I_{\rm{B}}}, \label{eq4}
\end{equation}
where ${{\boldsymbol{W}}_{c,(j,:)}} = [\begin{array}{*{20}{c}}
{{w_{c,(j,1)}}},&{{w_{c,(j,2)}}},& \ldots, &{{w_{c,(j,{N_{c,\rm{R}}})}}}
\end{array}] \in {\mathcal{R}^{1 \times {N_{c,\rm{R}}}}}$ is the \textit{j}th row of the precoding matrix of the $c$th cell, and ${{\boldsymbol{d}}_c} = [\begin{array}{*{20}{c}}
{{d_{c,1}}},&{{d_{c,2}}},&\ldots,&{{d_{c,{N_{c,\rm{R}}}}}{]^{\rm{T}}} \in {\mathcal{R}^{{N_{c,\rm{R}}} \times 1}}}\end{array}$ is the data vector of normalized pulse amplitude modulation (PAM) symbols for all the users in the \textit{c}th cell. Since ${x_{c,j}}$ must satisfy ${x_{c,j}} \in \left[ {{I_{\rm{l}}},{I_{\rm{h}}}} \right]$, where ${I_{\rm{l}}}$ and ${I_{\rm{h}}}$ are the lower and upper bounds of the dynamic range of LEDs, respectively, we have
\vspace{-0.cm}
\begin{equation}
{{{\left\| {{\boldsymbol{W}}_{c,(j,:)}} \right\|}_1} \le \Delta I}, \label{eeq5}
\end{equation}
where ${\Delta I = \min \left( {{I_{\rm{B}}}{\rm{ - }}{I_{\rm{l}}},{I_{\rm{h}}} - {I_{\rm{B}}}} \right)}$. In this work, ZF precoding is adopted, such that
\begin{equation}
\hspace{-0.03cm}{{\boldsymbol{W}}_c} = {\boldsymbol{H}}_c^\dag {\rm{diag}}\left\{ {{{\left[ {\begin{array}{*{20}{c}}
{\sqrt {{q_{c,1}}}},&{\sqrt {{q_{c,2}}}},& \ldots, &{\sqrt {{q_{{N_{\rm{c,R}}}}}}}
\end{array}} \right]}^{\rm{T}}}} \right\}, \label{eq10}
\end{equation}
where ${\sqrt {{q_{c,i}}}}$ is the equivalent channel gain of the \textit{i}th user. ${\boldsymbol{H}}_c^\dag$ is the generalized inverse matrix of ${\boldsymbol{H}}_c$, which has several matrix forms. In this work, we adopt pseudo-inverse ${\boldsymbol{H}}_c^\dag = {\boldsymbol{H}}_c^{\rm{T}}{\left( {{{\boldsymbol{H}}_c}{\boldsymbol{H}}_c^{\rm{T}}} \right)^{ - 1}}$\cite{pseudoinv}.

The signal received by user $\textit{i}$ in the \textit{c}th cell after removing the direct current (DC) bias by alternating current (AC) coupling can be expressed as \eqref{eq6}, which is shown at the top of this page, where $\gamma$ and $\zeta$ are the responsivity of the PD and the electrical-to-optical conversion coefficient, respectively. ${{\boldsymbol{H}}_{c,(i,:)}} = \left[ {\begin{array}{*{20}{c}}
{{h_{c,(i,1)}}},&{{h_{c,(i,2)}}},&\ldots,&{{h_{c,(i,{N_{c,\rm{T}}})}}}
\end{array}} \right] \in {\mathcal{R}^{1 \times {N_{c,\rm{T}}}}}$, $c \in \left\{ {1, \cdots ,{N_c}} \right\}$ is the channel matrix between $N_{c,\rm{T}}$ LEDs and user $\textit{i}$ in the \textit{c}th cell. ${{\boldsymbol{x}}_c} = [\begin{array}{*{20}{c}}
{{x_{c,1}}},&{{x_{c,2}}},&\ldots,&{{x_{c,{N_{c,\rm{T}}}}}{]^T}}
\end{array} \in {\mathcal{R}^{{N_{c,\rm{T}}}} \times 1}$, $c \in \left\{ {1, \cdots ,{N_c}} \right\}$ is the transmitted signal vector and ${\boldsymbol{I}}_{\rm{B}}^c = \left[ {\begin{array}{*{20}{c}}
{{I_{{\rm{B}}}}},&{{I_{{\rm{B}}}}},&\cdots,&{{I_{{\rm{B}}}}}
\end{array}} \right] \in {\mathcal{R}^{{N_{c,\rm{T}}} \times {\rm{1}}}}, c \in \left\{ {1, \cdots ,{N_c}} \right\}$ is the DC bias vector of the \textit{c}th cell. ${{\boldsymbol{H}}_{c,(i,:)}}{{\boldsymbol{W}}_{c,(:,i)}}{d_{c,i}}$ is the desired signal part. Define the set of indexes of users in cell $c$ as ${\mathcal{U}}_c$,  ${{\boldsymbol{H}}_{c,(i,:)}}\sum\limits_{k \in {{\mathcal{U}}_c},k \ne i}^{{N_{c,\rm{R}}}} {{{\boldsymbol{W}}_{c,(:,k)}}{d_{c,k}}}$ and $\sum\limits_{c' \ne c}^{{N_c}} {\sum\limits_{j \in {{\mathcal{U}}_{c'}}}^{{N_{c',\rm{R}}}} {{{\boldsymbol{H}}_{c',(i,:)}}{{\boldsymbol{W}}_{c',(:,j)}}{d_{c',j}}} }$ are the intra and inter-CI of the \textit{c}th cell, respectively. The intra-CI can be eliminated by ZF precoding \cite{ZFdelMUI}. Besides, ${n_{c,i}}$ is the additive white Gaussian noise (AWGN) with zero mean and variance $\sigma _{c,i}^2$, which is written as\cite{ratemaxZF}
\setcounter{equation}{5}
\begin{equation}
\sigma _{c,i}^2 = 2\gamma{e_{\rm{ch}}}\overline {P_r^{c,i}} B + 4\pi {e_{\rm{ch}}}{A_r}\gamma {\chi _{amp}}\left( {1 - \cos \left( \Psi  \right)} \right)B + i_{amp}^2B, \label{eq7}
\end{equation}
where ${e_{\rm{ch}}}$ is the elementary charge, $B$ is the system bandwidth, ${i_{amp}}$ is the pre-amplifier noise current density, $\overline {P_r^{c,i}} {\rm{ = }}E{\rm{[}}P_r^{c,i}] = \zeta \left( {{{\boldsymbol{H}}_{c,i,:}}{\boldsymbol{I}}_{\rm{B}}^c} \right)$ is the average received optical power of user $i$ in the \textit{c}th cell, and ${\chi _{amp}}$ is the ambient light photocurrent.
\vspace{-0.cm}
\subsection{Channel Capacity bound}
\vspace{-0.cm}
Next, we introduce a sum-rate maximization problem for hybrid dimming scheme in MU-MC-MISO VLC systems. We first derive a closed-form expression for the achievable sum-rate of the MU-MISO system. In particular, the channel capacity of the $i$th user in the \textit{c}th cell is lower bounded by \cite{infoTheo}

\vspace{-0.cm}
\begin{align}
{C_{c,k}} &= I\left( {{X_{c,i}};{Y_{c,i}}} \right)\notag\\
 &= h\left( {{Y_{c,i}}} \right) - h\left( {{Y_{c,i}}\left| {{X_{c,i}}} \right.} \right)\notag\\
 &= h\left( {{X_{c,i}} + {T_{c,i}}} \right) - h\left( {{T_{c,i}}} \right)\notag\\
 &\ge \frac{1}{2}{\log _{\rm{2}}}\left( {{e^{2h\left( {{X_{c,i}}} \right)}} + {e^{2h\left( {{T_{c,i}}} \right)}}} \right) - h\left( {{T_{c,i}}} \right)\notag\\
 &= \frac{1}{2}{\log _{\rm{2}}}\left( {1 + \frac{{{e^{2h\left( {{X_{c,i}}} \right)}}}}{{{e^{2h\left( {{T_{c,i}}} \right)}}}}}\right),\label{capa1}
 \end{align}where $h\left( \cdot  \right)$ is the entropy function. ${X_{c,i}}$, ${T_{c,i}}$, and ${Y_{c,i}}$ denote the random variables corresponding to the desired signal ${\overline x _{c,i}} = \gamma \varsigma {{\boldsymbol{H}}_{c,(i,:)}}{{\boldsymbol{W}}_{c,(:,i)}}{d_{c,i}}$, the sum of interference and noise ${t_{c,i}} = \gamma \varsigma \sum\limits_{c' \ne c}^{{N_c}} {\sum\limits_{j \in {{\mathcal{U}}_{c'}}}^{{N_{c',\rm{R}}}} {{{\boldsymbol{H}}_{c',(i,:)}}{{\boldsymbol{W}}_{c',(:,j)}}{d_{c',j}}} }  + {n_{c,i}}$ and the received signal ${y_{c,i}}$, respectively. Since ${d_{c,i}}$ has zero mean and is normalized to the range of $[-1, 1]$, we have $h\left( {{X_{c,i}}} \right) = \log \left( {2\gamma \varsigma {{\boldsymbol{H}}_{c,(i,:)}}{{\boldsymbol{W}}_{c,(:,i)}}} \right)$. $h\left( {{T_{c,i}}} \right)$ is upper bounded by the differential entropy of a Gaussian random variable with variance $ \sigma _{{t_{c,i}}}$, written as $h\left( {{T_{c,i}}} \right) \le \frac{1}{2}\log \left( {2\pi e\sigma _{{t_{c,i}}}^2} \right)$, where $\sigma _{{t_{c,i}}}^2 = {\left( {\gamma \varsigma } \right)^2}\sum\limits_{c' \ne c}^{{N_c}} {\sum\limits_{j \in {{\mathcal{U}}_{c'}}}^{{N_{c',\rm{R}}}} {\sigma _{{d_{c,i}}}^2{{\left( {{{\boldsymbol{H}}_{c',(i,:)}}{{\boldsymbol{W}}_{c',(:,j)}}} \right)}^2} + \sigma _{c,i}^2} }$, and $\sigma _{{d_{c,i}}}^2 = \frac{1}{3}$ denotes the variance of the transmit symbol ${d_{c,i}}$, since ${d_{c,i}}$ obeys the uniform distribution. Therefore, the lower bound of ${C_{c,i}}$ is
\vspace{-0.cm}
 \begin{equation}
C_{c,i}^{\rm{L}}{\rm{ = }}\frac{1}{2}\log \left( {1 + {\xi _{c,i}}} \right) \buildrel \Delta \over = {R_{c,i}}, \label{capa2}
\end{equation}
where ${\xi _{c,i}}$ is the signal to interference and noise ratio (SINR) of the $i$th user in the $c$th cell, written as \cite{SINR13}
\begin{equation}
\hspace{-0.cm}{\xi _{c,i}} = \frac{{2{{\left( {\gamma \varsigma } \right)}^2}{{\left| {{{\boldsymbol{H}}_{c,(i,:)}}{{\boldsymbol{W}}_{c,(:,i)}}} \right|}^2}}}{{\pi e\left( {\frac{{{{\left( {\gamma \varsigma } \right)}^2}}}{3}\sum\limits_{c' \ne c}^{{N_c}} {\sum\limits_{j \in {{\mathcal{U}}_{c'}}}^{{N_{c',r}}} {{{\left( {{{\boldsymbol{H}}_{c',(i,:)}}{{\boldsymbol{W}}_{c',(:,j)}}} \right)}^2} + \sigma _{c,i}^2} } } \right)}}. \label{sinr}
\end{equation}
Then, define the LED selection matrix of the $c$th cell ${{\boldsymbol{A}}_c} = [\begin{array}{*{20}{c}}{{\boldsymbol{A}}_{c,1}},&{{\boldsymbol{A}}_{c,2}},& \ldots, &{{\boldsymbol{A}}_{c,{N_{c,\rm{T}}}}}\end{array}] \in {\mathcal{R}^{{N_{c,\rm{T}}} \times {N_{c,\rm{T}}}}}$, which includes ${N_{c,\rm{T}}}$ column vectors ${\boldsymbol{A}}_{c,j} \in {\mathcal{R}^{{N_{c,\rm{T}}} \times 1}}, \forall j \in \left\{ {1,2, \cdots ,{N_{c,\rm{T}}}} \right\}$. If the $\textit{j}$th LED is selected, ${{\boldsymbol{A}}_{c,j}} = {\bm{e}_j} \in {\mathcal{R}^{{N_{c,{\rm{T}}}} \times 1}}$ is a unit vector with the $j$th entry being 1 and the others being 0s; otherwise, ${\boldsymbol{A}}_{c,j}={\boldsymbol{0}}$. With the LED selection, the SINR is written as
\vspace{-0.cm}
\begin{equation}
\hspace{-0.3cm} {\xi _{c,i}} = \frac{{2{{\left( {\gamma \varsigma } \right)}^2}{{\left| {{{\boldsymbol{H}}_{c,(i,:)}}{{\boldsymbol{A}}_c}{{\boldsymbol{W}}_{c,(:,i)}}} \right|}^2}}}{{\pi e\left( {\frac{{{{\left( {\gamma \varsigma } \right)}^2}}}{3}\sum\limits_{c' \ne c}^{{N_c}} {\sum\limits_{j \in {{\mathcal{U}}_{c'}}}^{{N_{c',{\rm{R}}}}} {{{\left( {{{\boldsymbol{H}}_{c',(i,:)}}{{\boldsymbol{A}}_{c'}}{{\boldsymbol{W}}_{c',(:,j)}}} \right)}^2} + \sigma _{c,i}^2} } } \right)}}. \label{sinrA}
\end{equation}
With \eqref{sinrA}, the sum-rate of the system can be written as

\vspace{-0.cm}
\begin{equation}
R = \frac{1}{2}\sum\limits_{c = 1}^{{N_c}} {\sum\limits_{i = 1}^{{N_{c,{\rm{R}}}}} {\log \left( {1 + {\xi _{c,i}}} \right)}}. \label{sumrateA}
\end{equation}
From \eqref{sinrA} we can observe that the sum-rate is largely dependent on the inter-CI $\sum\limits_{c' \ne c}^{{N_c}} {\sum\limits_{j \in {{\mathcal{U}}_{c'}}}^{{N_{c',\rm{R}}}} {{{\left( {{{\boldsymbol{H}}_{c',(i,:)}}{{\boldsymbol{A}}_{c'}}{{\boldsymbol{W}}_{c',(:,j)}}} \right)}^2}} }$. In order to alleviate inter-CI while satisfying the illumination uniformity constraint, both the LED selection and the precoding design should be jointly considered. This strategy fortunately coincides with the principle of hybrid dimming, which will be introduced next.

\vspace{-0.cm}
\subsection{Hybrid Dimming Scheme}
This subsection specifies the hybrid dimming control scheme, which combines SD and AD. Denote ${\boldsymbol{A}} = {\rm{diag}}\left\{ {{{\boldsymbol{A}}_1},{{\boldsymbol{A}}_2}, \cdots ,{{\boldsymbol{A}}_{{N_c}}}} \right\} \in {\mathcal{R}^{{N_{\rm{T}}} \times {N_{\rm{T}}}}}$ as the combination form of the LED selection matrices of all $N_c$ cells. The dimming level is defined as
\vspace{-0.cm}
\begin{equation}
\eta {\rm{ = }}\frac{{{\left\| {{\boldsymbol{A}}} \right\|}_1}{({I_{\rm{B}}} - {I_{\rm{l}}})}}{{{N_{\rm{T}}}({I_0} - {I_{\rm{l}}})}} \times 100\%,\label{illum}
\end{equation}
where ${{\left\| {{\boldsymbol{A}}} \right\|}_1}={n_t}$ is the number of activated LEDs and ${I_0} = {{\left( {{I_{\rm{l}}} + {I_{\rm{h}}}} \right)} \mathord{\left/
 {\vphantom {{\left( {{I_{\rm{l}}} + {I_{\rm{h}}}} \right)} 2}} \right.
 \kern-\nulldelimiterspace} 2}$.
The signal beyond the dynamic current range of LEDs $\left[ {{I_{\rm{l}}},{I_{\rm{h}}}} \right]$ has to be clipped, since it results in clipping noise\cite{clipnoise}. Therefore, the VLC signals need to be within the dynamic range of LEDs. However, the value of ${I_{\rm{B}}}$ can significantly affect the VLC signal range. For instance, ${I_{\rm{B}}}$ may be high to satisfy a high dimming level requirement, which results in clipping noise at the upper bound of the dynamic range of LEDs. To avoid such side effects, we propose a two-step dimming method. In the first step, the number of activated LEDs $n_t$ is adjusted to achieve coarse dimming control. Then the DC-bias level is adjusted to achieve precise dimming control. In particular, we first round down the number of activated LEDs as
\vspace{-0.cm}
\begin{equation}
{n_t} = \left\lfloor {\eta {N_{\rm{T}}}} \right\rfloor. \label{nt}
\end{equation}
Then the DC bias can be obtained as:
\begin{equation}
{I_{\rm{B}}} = \frac{{\eta {N_{\rm{T}}}\left( {{I_0} - {I_{\rm{l}}}} \right)}}{{{n_t}}} + {I_{\rm{l}}}. \label{IB}
\end{equation}

Furthermore, the coefficient of variation of root mean square error (CV(RMSE)) is used to quantify the illumination uniformity \cite{cvrmse}, which is defined as
\begin{equation}
{\rm{CV(RMSE) = }}\frac{\upsilon}{\overline E }, \label{CVrmse}
\end{equation}
where $\upsilon$ is the root mean square error of illumination, and $\overline E$ is the average illumination. Define $K$ as the total number of the sample points on the receiver plane, ${{\boldsymbol{E}}_\mu } = [\begin{array}{*{20}{c}}
{{E_{\mu ,1}}}&{{E_{\mu ,2}}}& \cdots &{{E_{\mu ,{N_{\rm{T}}}}}}
\end{array}] \in {\mathcal{R}^{1 \times {N_{\rm{T}}}}}$ as the illuminance vector of the $\mu$th sample point, then $\overline E$ is given by
\begin{equation}
\overline E \left( {{{\bf{E}}_\mu },{\boldsymbol{A}}} \right) = \mathop {{\rm{avg}}}\limits_{\mu  \in \left\{ {1,2, \cdots ,K} \right\}} \left\{ {{{\left\| {{{\bf{E}}_\mu }{\boldsymbol{A}}} \right\|}_1}} \right\}.\label{Eavg}
\end{equation}
In addition, the illumination root mean square error at the receiver plane can be expressed as
\begin{equation}
\upsilon  = \sqrt {\frac{1}{K}\sum\limits_{\mu  = 1}^K {{{\left( {{{\left\| {{{\boldsymbol{E}}_\mu }{\boldsymbol{A}}} \right\|}_1} - {\overline E}} \right)}^2}} } .\label{rmse}
\end{equation}
The horizontal illuminace in lux of the $j$th LED received at the $\mu$th sample point ${E_{\mu ,j}},\forall j \in \left\{ {1,2, \cdots ,{N_{\rm{T}}}} \right\}$ can be represented as \cite{horillum}:
\begin{equation}
{E_{\mu ,j}} = {{I(0) \times {{\cos }^l}{\phi _{\mu ,j}}\cos {\psi _{\mu ,j}}} \mathord{\left/
 {\vphantom {{I(0) \cdot {{\cos }^l}{\phi _{\mu ,j}}\cos {\psi _{\mu ,j}}} {d_{\mu ,j}^2}}} \right.
 \kern-\nulldelimiterspace} {d_{\mu ,j}^2}}, \label{ilumnance}
\end{equation}
where ${I(0)}$ is the maximum luminous intensity, ${\phi _{\mu ,j}}$, ${\psi _{\mu ,j}}$ and $d_{\mu ,j}$ are the angle of irradiance, the angle of incidence, and the distance between the $j$th LED and the $\mu$th point on the receiver plane, respectively.

\subsection{Problem Formulation}
With the illumination constraints, the target of this work is to maximize the sum-rate of users by jointly optimizing the LED selection and precoding matrix. The optimization problem is formulated as

\vspace{-0.4cm}
\begin{align}
&\hspace{-0.3cm}\mathop {\max }\limits_{\scriptstyle{{\boldsymbol{W}}_c},{{\boldsymbol{A}}_c}} R\left( {{{\boldsymbol{W}}_c},{{\boldsymbol{A}}_c}} \right),\label{maxsumrate} \\
&\hspace{-0.2cm}{\rm{s.t.}} \quad {{\boldsymbol{H}}_c}{{\boldsymbol{A}}_c}{{\boldsymbol{W}}_c} = {\rm{diag}}\{ \sqrt{{\boldsymbol{q}}_c}\},\tag{\ref{maxsumrate}{a}} \label{maxsumratea} \\
&\quad\begin{array}{*{20}{c}}
{{{\left\| {{{\left[ {{{\boldsymbol{A}}_c}{{\boldsymbol{W}}_c}} \right]}_{j,:}}} \right\|}_1} \le \Delta I},&{\forall j \in \left\{ {1,2, \cdots ,{N_{\rm{c,T}}}} \right\}},
\end{array}\tag{\ref{maxsumrate}{b}} \label{maxsumrateb}  \\
&\quad\quad {\rm{CV(RMSE)}} \le {U_{{\rm{th}}}}, \tag{\ref{maxsumrate}{c}} \label{maxsumratec} \\
&\quad\quad \eta {\rm{ = }}\frac{{{{{\left\| {\boldsymbol{A}} \right\|}_1}}({I_{\rm{B}}} - {I_{\rm{l}}})}}{{{N_{\rm{T}}}({I_0} - {I_{\rm{l}}})}} \times 100\%, \tag{\ref{maxsumrate}{d}}  \label{maxsumrated}\\
&\quad\begin{array}{*{20}{c}}
{{q_{c,i}} > 0},&{\forall i \in \left\{ {1,2, \cdots ,{N_{\rm{c,R}}}} \right\}},
\end{array} \tag{\ref{maxsumrate}{e}} \label{maxsumratee}\\
&\quad\begin{array}{*{20}{c}}
{a_{c,j}} \in \left\{ {0,1} \right\},&{\forall j \in \left\{ {1,2, \cdots ,{N_{\rm{c,T}}}} \right\}},
\end{array}\tag{\ref{maxsumrate}{f}}\label{maxsumratef}
\end{align}
where ${\sqrt{{\boldsymbol{q}}_c}} = {\left[ {\begin{array}{*{20}{c}}
{\sqrt{q_{c,1}}},&{\sqrt{q_{c,2}}},&\cdots,&{\sqrt{q_{c,{N_{c,\rm{R}}}}}}
\end{array}} \right]^{\rm{T}}} \in {\mathcal{R}^{{N_{c,\rm{R}}} \times 1}}$, ${U_{\rm{th}}}$ is the threshold of illumination uniformity, and ${a_{c,j}}$,${\forall j \in \left\{ {1,2, \cdots ,{N_{c,\rm{T}}}} \right\}}$ is a diagonal element of ${{\boldsymbol{A}}_c}$. In \eqref{maxsumrate}, the optimization variables are the precoding matrix ${{\boldsymbol{W}}_c}$ and the LED selection matrix ${{\boldsymbol{A}}_c},\forall c = \left\{ {1,2, \cdots ,{N_c}} \right\}$. The objective function \eqref{maxsumrate} is the achievable sum-rate of users according to \eqref{sumrateA}. \eqref{maxsumratea} is the ZF constraint, which implies that the channel matrix after precoding is a diagonal matrix. \eqref{maxsumrateb} implies that the amplitude of the precoding matrix must be in the range of $\Delta I$ to satisfy ${x_{c,j}} \in \left[ {{I_{\rm{l}}},{I_{\rm{h}}}} \right]$. \eqref{maxsumratec} and \eqref{maxsumrated} are the illumination uniformity constraint and illumination level constraint, respectively.  \eqref{maxsumratee} indicates that the elements of ${{\boldsymbol{q}}_c}$ must be positive. \eqref{maxsumratef} indicates that the diagonal elements of ${{\boldsymbol{A}}_c}$ are 0-1 integers.

\vspace{0.cm}
\section{Joint Design of LED Selection And Precoding Matrix}
In this section, we solve \eqref{maxsumrate} by jointly design of LED selection and precoding matrix. First, the cell formation should be specified. In this work, the cells are initially formed by the distance-based user-centric (UC) amorphous cell formation \cite{MC1} and adjusted with different activated LEDs patterns under varied dimming levels. Then we propose an efficient two-step algorithm. Since ${{\boldsymbol{A}}_c}$ is a Boolean matrix and the objective function is non-convex, \eqref{maxsumrate} is a non-convex mixed integer problem \cite{ncmip} and its direct solution is computationally intractable. Thus, we separate the original problem into two subproblems, which will be explained in the following two sections. In the first subproblem, we optimize only the LED selection matrix ${{\boldsymbol{A}}_c}$ with a fixed value of ${{\boldsymbol{W}}_c}$ and adjust the cell formation. Then, in the second subproblem, we obtain the optimal precoding matrix ${{\boldsymbol{W}}_c}$ with maximum sum-rate of users based on ${{\boldsymbol{A}}_c}$ calculated in the previous step. These two subproblems are solved iteratively until the original objective function converges.
\vspace{-0.cm}
\subsection{LED Selection with Cell Formation}
\begin{figure}[t!]
	\subfigure[\small $d_0$=3 m]{
		\begin{minipage}{1\linewidth}
			\centerline{\includegraphics[width=0.7\linewidth]{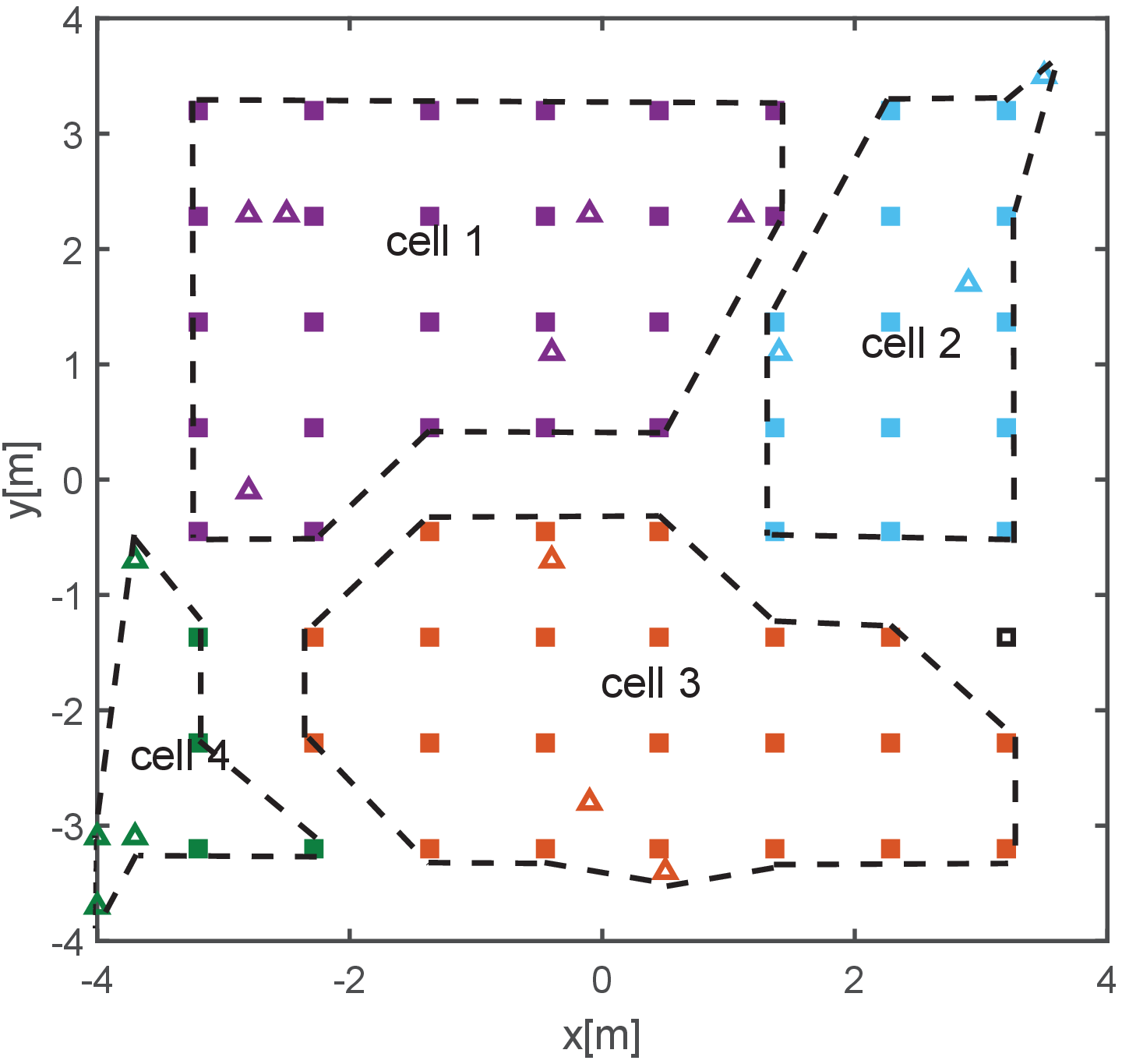}}
		\end{minipage}%
	}%
	
	\subfigure[\small $d_0$=2.5 m]{
		\begin{minipage}{1\linewidth}
			\centerline{\includegraphics[width=0.7\linewidth]{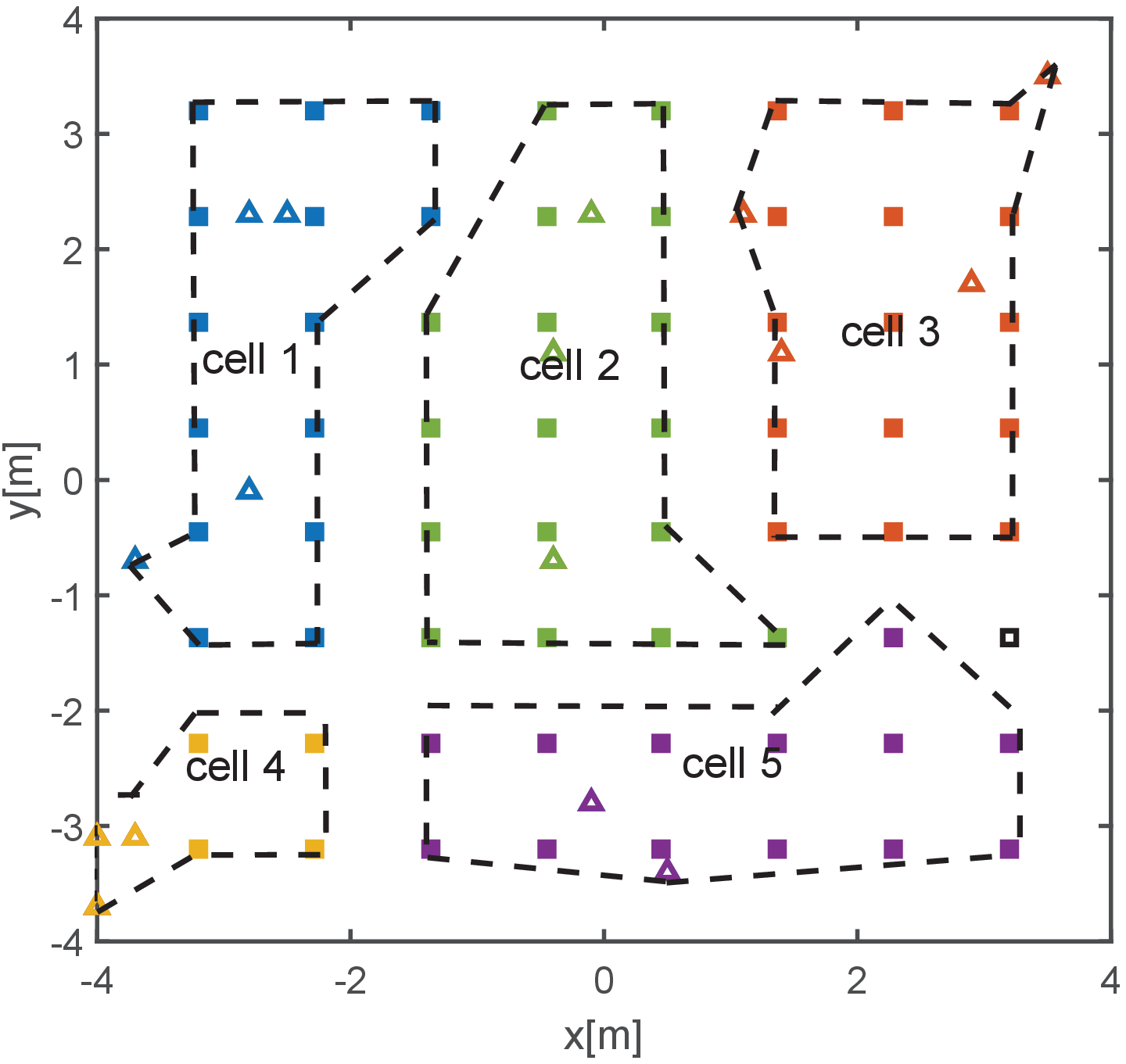}}
		\end{minipage}%
	}%
	\vspace{-0.3cm}
	\caption{UC amorphous cell formation of $N_{\rm{T}}=64$ with different distance thresholds $d_0$.}\label{fig:figcell}
	\vspace{-0.cm}
\end{figure}

This subsection introduces the first subproblem in our proposed two-step algorithm. The first subproblem optimizes the LED selection matrix to maximize the sum-rate with fixed precoding matrix and adjusts the cell formation.

Before solving this subproblem, the cell formation should be initialized. Note that the proposed TASP-HD can be applied for any cell formation, and the UC cell formation is adopted here due to its high energy efficiency. The adopted UC cell formation initialization algorithm \cite{MC1} is described briefly as follows:

\textit{1. Distance-based User clustering}: Denote the cluster of all the ${N_{\rm{R}}}$ users as $\mathcal{U}$ and the cluster of users in the \textit{c}th cell as $\mathcal{U}_c$. In this step, users in $\mathcal{U}$ are assigned into $\mathcal{U}_1,\mathcal{U}_2,\cdots,\mathcal{U}_{N_c}$ according to a pre-defined distance threshold ${d_0}$. In particular, the users are clustered by following steps:

1) Initialize the counter $c=1$, $\mathcal{U}_1,\mathcal{U}_2,\cdots,\mathcal{U}_{N_c}=\emptyset $.

2) Initialize $\mathcal{U}_c$ by recruiting the first user who has not been included in any clusters and using the location of this user as the centroid of $\mathcal{U}_c$.

3)	Recruit another user into $\mathcal{U}_c$ who has not been assigned to any clusters and has the distance to the centroid of $\mathcal{U}_c$ shorter than ${d_0}$. Then update the centroid as the geometric center of all the users in $\mathcal{U}_c$. Repeat this step until no other user can be added into $\mathcal{U}_c$.

4) Set $c=c+1$ and repeat Step 2), 3). Finally, all the users are allocated into a cluster, such that $\mathcal{U} = {\mathcal{U}_1} \cup {\mathcal{U}_2}\cup\cdots {{\mathcal{U}}_{{N_c}}}$.

\textit{2. LED association}: In this step, LEDs make association to users based on the channel gains. Denote the set of all ${N_{\rm{T}}}$ LEDs as $\mathcal{T}$, and the set of LEDs in the \textit{c}th cell as $\mathcal{T}_c$. Since some of the LEDs may have no LOS links with any users due to the FOV constraint, we construct association matrix ${\boldsymbol{M}} \in {\mathcal{R}^{{N_{\rm{R}}} \times {N_{\rm{T,LOS}}}}}$, where ${N_{\rm{T,LOS}}}$ is the number of LEDs having LOS links to users. The initial value of ${\boldsymbol{M}}$ is the channel gains, written as ${\boldsymbol{M}} = {{\boldsymbol{H}}_{\rm{LOS}}}$. Then the steps of LED association are as follows:

1) \textit{One LED to one user association}: For each user, find the best user-LED association [${i}$, ${j}$*] with the strongest LOS channel amongst the $\textit{i}$th rows of ${\boldsymbol{M}}$. Then set the $\textit{j*}$th column of ${\boldsymbol{M}}$ to ${\boldsymbol{0}}$. Repeat this step until all the users find the best matched LEDs.

2) \textit{Multiple LEDs to one user association}: For the remaining non-zero $({N_{\rm{T,LOS}}} - {N_{\rm{R}}})$ columns of ${\boldsymbol{M}}$, find the strongest LOS channel gain amongst each column, then set the column to ${\boldsymbol{0}}$. For example, if the $\textit{n}$th entry of the $\textit{m}$th column is the strongest LOS channel, then the $\textit{m}$th LED is allocated to the $\textit{n}$th user, written as user-LED pair [${n}$, ${m}$].

Finally ${\boldsymbol{M}}={\boldsymbol{0}}$ and all the ${N_{\rm{T,LOS}}}$ LEDs are allocated to ${N_{\rm{R}}}$ users. The LEDs are allocated into $\mathcal{T}_c$ if their associated users are in $\mathcal{U}_c$. The $\textit{c}$th cell is thus composed of $\left\{ {{\mathcal{U}_c},{\mathcal{T}_c}} \right\}$ and each user must have at least one associated LED.
The number and the size of the amorphous cells change with different distance thresholds and values of FOV. Figure \ref{fig:figcell} illustrates examples of different cell formations in a 8 m$\times$8 m$\times$2.5 m square room with ${N_{\rm{R}}}=16$ (marked by triangles), ${N_{\rm{T}}}=64$ (marked by squares), and the FOV of PDs is ${60^ \circ }$. The distance thresholds $d_0$ are set to 3 m and 2.5 m, respectively.

With initialized UC cell formation, the first subproblem to optimize the LED selection matrix ${{\boldsymbol{A}}_c}$ is given by
\begin{align}
\hspace{-0mm}
&\mathop {\max }\limits_{{{\boldsymbol{A}}_c}}\quad\quad\quad R\left( {{{\boldsymbol{A}}_c}} \right),\label{sub1} \\
\hspace{-2.7mm}
&\rm{s.t.}\quad \eqref{maxsumratec},\eqref{maxsumrated},\eqref{maxsumratef}. \notag
\end{align}
Since ${{\boldsymbol{A}}_c} \in {\mathcal{R}^{{N_{{\rm{c,T}}}} \times {N_{{\rm{c,T}}}}}},c \in \{ 1,2, \cdots ,{N_c}\}$ are diagonal matrices whose diagonal entries vectors are ${{\boldsymbol{a}}_c} = {\left\{ {0,1} \right\}^{1\times{N_{c,{\rm{T}}}}}}$, this problem is a mixed integer non-linear programming (MINLP) problem, and thus it is NP-hard\cite{exhau}. The NP-hard problem can be solved by exhaustive search for all possible values of ${{\boldsymbol{a}}_c}$ whose computational complexity is exponentially increased with ${N_{\rm{T}}}$, which will be extremely high with a large size of ${{\boldsymbol{A}}_c}$ \cite{exhau}. The branch-and-bound (B\&B) algorithm is another widely adopted algorithm to solve MINLP with lower computational complexity than exhaustive search\cite{ncmip}. B\&B algorithm solves problem iteratively, and each iteration has two branches. The optimization problem needs to be solved on each branch. Hence, the computation cost of B\&B algorithm is still high\cite{penalty}. This motivates us to propose another simple method by relaxing the integer variables ${{\boldsymbol{a}}_c}=\left[ {\begin{array}{*{20}{c}}
		{{a_{c,1}},}&{{a_{c,2}}},& \cdots &{{a_{c,{N_{c,{\rm{T}}}}}}}
\end{array}} \right]$ into continuous ones in $\left[ {{\rm{0}},{\rm{1}}} \right]$ and reformulate problem \eqref{sub1} as follows.

\noindent
\textbf{Proposition 1}: Given a sufficiently large coefficient $\lambda\to \infty$ and continuous variables ${a_{c,j}}$ in $\left[ {{\rm{0}},{\rm{1}}} \right]$, $\forall c \in \{ 1,2, \cdots ,{N_c}\}, \forall j \in \left\{ {1,2, \cdots ,{N_{c,{\rm{T}}}}} \right\}$, \eqref{sub1} is equivalent to the following problem
\vspace{-0.cm}
\begin{align}
\hspace{-0mm}
&\mathop {\max }\limits_{{a_{c,j}} \in \left[ {{\rm{0}},{\rm{1}}} \right]} R(a_{c,j}) - \lambda \sum\limits_{c = 1}^{{N_c}} {\sum\limits_{j = 1}^{{N_{c,\rm{T}}}} {\left( {{a_{c,j}} - a_{c,j}^2} \right),} }\label{provequal} \\
\hspace{-0.mm}
&\rm{s.t.}\quad\quad\quad\quad\quad \eqref{maxsumratec},\eqref{maxsumrated}.\notag
\end{align}
\textbf{Proof}: To prove that \eqref{sub1} and \eqref{provequal} are equal, we only need to prove that $\mathop {\max }\limits_{{a_{c,j}} \in \left[ {{\rm{0}},{\rm{1}}} \right]} R(a_{c,j}) - \lambda \sum\limits_{c = 1}^{{N_c}} {\sum\limits_{j = 1}^{{N_{c,\rm{T}}}} {\left( {{a_{c,j}} - a_{c,j}^2} \right)} }$ and $\mathop {\max }\limits_{{a_{c,j}} \in \left\{ {0,1} \right\}} R(a_{c,j}) \ $ share the same optimal solution. To this end, we first relax the integer $a_{c,j}$ into continuous variables. In particular, define the set $\mathcal{D}$ as
\begin{small}
\begin{equation}
\hspace{-0.cm} \mathcal{D} = \left\{ {{{\boldsymbol{a}}_1}, \cdots ,{{\boldsymbol{a}}_{{N_c}}} \in {{\left[ {0,1} \right]}^{{N_{c,{\rm{T}}}}}}|\sum\limits_{c = 1}^{{N_c}} {\sum\limits_{j = 1}^{{N_{c,{\rm{T}}}}} { \left({a_{c,j}} - a_{c,j}^2\right) \le 0} } } \right\}.
\end{equation}
\end{small}Obviously the set $\left\{ {{{\boldsymbol{a}}_c}\left| {{{\boldsymbol{a}}_c} \in {{\left\{ {0,1} \right\}}^{{N_{c,{\rm{T}}}}}}}, c \in \left\{ {1, \cdots ,{N_c}} \right\} \right.} \right\}$ is equivalent to  $\mathcal{D}$. Therefore $\mathop {\max }\limits_{{a_{c,j}} \in \left\{ {0,1} \right\}} R(a_{c,j}) \ $ is equivalent to
\begin{equation}\label{maxRequal}
\begin{split}
&\mathop {\max }\limits_{{a_{c,j}} \in \left[ {0,1} \right]}\quad\quad R(a_{c,j}) \\
& {\rm{s.t.}} \sum\limits_{c = 1}^{{N_c}} {\sum\limits_{j = 1}^{{N_{c,{\rm{T}}}}} {\left({a_{c,j}} - a_{c,j}^2\right) \le 0}. }
\end{split}
\end{equation}

Next we only need to prove that $\mathop {\max }\limits_{{a_{c,j}} \in \left[ {{\rm{0}},{\rm{1}}} \right]} R(a_{c,j}) - \lambda \sum\limits_{c = 1}^{{N_c}} {\sum\limits_{j = 1}^{{N_{c,\rm{T}}}} {\left( {{a_{c,j}} - a_{c,j}^2} \right)} }$ shares the same optimal solution with \eqref{maxRequal}, which can also be expressed as
\begin{equation}\label{maxRequalmin}
\begin{split}	
&\mathop {\min }\limits_{{a_{c,j}} \in \left[ {0,1} \right]}\quad\quad - R(a_{c,j})\\
&{\rm{s.t.}} \sum\limits_{c = 1}^{{N_c}} {\sum\limits_{j = 1}^{{N_{c,{\rm{T}}}}} {\left({a_{c,j}} - a_{c,j}^2 \right) \le 0} .}
\end{split}
\end{equation}
Suppose the solution of $\mathop {\max }\limits_{{a_{c,j}} \in \left\{ {0,1} \right\}} R(a_{c,j}) \ $ exists and the optimal value is denoted by ${R^*}$. Hence, the optimal value of \eqref{maxRequalmin} is denoted as $(-{R^*})$. The Lagrangian of \eqref{maxRequalmin} can be written as $L({a_{c,j}},\lambda ) = -R(a_{c,j}) + \lambda \sum\limits_{c = 1}^{{N_c}} {\sum\limits_{j = 1}^{{N_{c,\rm{T}}}} {\left( {{a_{c,j}} - a_{c,j}^2} \right)} }$ with dual variable $\lambda > 0$. Denote $\chi \left( \lambda  \right)$ as the optimal value of the Lagrange dual function $\mathop {\min }\limits_{{a_{c,j}} \in \left[ {0,1} \right]} L\left( {{a_{c,j}},\lambda } \right)$, and denote ${a_{c,j \to \lambda }}$ as the value of ${a_{c,j}}$ with a given $\lambda$. Due to the weak duality property, the optimal solution of the Lagrange dual function must satisfy
\begin{equation}
\mathop {\max }\limits_{\lambda  \ge 0} \chi \left( \lambda  \right) \le (-{R^*}). \label{lagproof1}
\end{equation}
Note that ${a_{c,j}} - a_{c,j}^2 \ge 0$ when ${a_{c,j}} \in \left[ {{\rm{0}},{\rm{1}}} \right]$. Hence $\chi \left( \lambda  \right)$ increases with $\lambda$ when ${a_{c,j}} \in \left[ {{\rm{0}},{\rm{1}}} \right]$ and is upper bounded by $(-{R^*})$. Next, we show that $\mathop {\max }\limits_{\lambda  \ge 0} \chi \left( \lambda  \right) = (-{R^*})$ always holds. In the first case, if there exists ${a_{c,j \to {\lambda _0}}}$ satisfying $\sum\limits_{c = 1}^{{N_c}} {\sum\limits_{j = 1}^{{N_{c,{\rm{T}}}}} {\left( {{a_{c,j \to {\lambda _0}}} - {a^2}_{c,j \to {\lambda _0}}} \right)} }  = 0$, then ${\left( {a_{c,j \to {\lambda _0}}},{\lambda _0}\right)}$ is feasible for \eqref{maxRequalmin}, so
\begin{equation}
\chi \left( {\lambda _0}  \right) \ge (-{R^*}). \label{lagproof2}
\end{equation}
Combining \eqref{lagproof1} and \eqref{lagproof2}, we have $\chi \left( {\lambda _0}  \right) = -{R^*}$.

In the second case, if there does not exist  ${a_{c,j \to {\lambda _0}}}$ satisfying $\sum\limits_{c = 1}^{{N_c}} {\sum\limits_{j = 1}^{{N_{c,{\rm{T}}}}} {\left( {{a_{c,j \to {\lambda _0}}} - {a^2}_{c,j \to {\lambda _0}}} \right)} }  = 0$, we have ${a_{c,j \to \lambda }} - {a^2}_{c,j \to \lambda } > 0$ for all $\lambda  > 0$. When $\lambda \to \infty$, $\chi \left( {\lambda}  \right) \to \infty$ implies that $(-{R^*})\to \infty$, and thus $\mathop {\max }\limits_{{a_{c,j}} \in \left\{ {0,1} \right\}} R(a_{c,j}) \ $ has no solutions. This is contradictory to the assumption that $\mathop {\max }\limits_{{a_{c,j}} \in \left\{ {0,1} \right\}} R(a_{c,j}) \ $ has a solution. Therefore, ${a_{c,j \to \infty }} - {a^2}_{c,j \to \infty } = 0$ must hold when $\lambda \to \infty$. This means that the optimal value of $L\left( {{a_{c,j}},\lambda } \right)$ is obtained with ${a_{c,j \to \infty }}$, so we have
\begin{equation}
\chi \left( \infty  \right) \ge (-{R^*}). \label{lagproof3}
\end{equation}
From \eqref{lagproof1} and \eqref{lagproof3}, we have $\mathop {\max }\limits_{\lambda  \ge 0} \chi \left( \lambda  \right) = \chi \left( \infty  \right) = - {R^*}$. Therefore, we can easily derive that the optimal value of $-L({a_{c,j}},\lambda ) = R(a_{c,j}) - \lambda \sum\limits_{c = 1}^{{N_c}} {\sum\limits_{j = 1}^{{N_{c,\rm{T}}}} {\left( {{a_{c,j}} - a_{c,j}^2} \right)} }$ is $ {R^*}$, which means $\mathop {\max }\limits_{{a_{c,j}} \in \left\{ {0,1} \right\}} R(a_{c,j}) \ $ and $\mathop {\max }\limits_{{a_{c,j}} \in \left[ {{\rm{0}},{\rm{1}}} \right]} R(a_{c,j}) - \lambda \sum\limits_{c = 1}^{{N_c}} {\sum\limits_{j = 1}^{{N_{c,\rm{T}}}} {\left( {{a_{c,j}} - a_{c,j}^2} \right)} }$ share the same optimal value when $\lambda \to \infty$. The proof is ended.\hfill $\blacksquare$

Then we construct a exterior penalty function by \textbf{proposition 1}, written as \eqref{provequal}, where $\lambda \gg {\rm{1}}$ is a large constant acting as a penalty factor, and $ - \lambda \sum\limits_{c = 1}^{{N_c}} {\sum\limits_{j = 1}^{{N_{c,\rm{T}}}} {\left( {{a_{c,j}} - a_{c,j}^2} \right)} }$ is a penalty term which penalizes the objective function for any value of ${a_{c,j}}$ other than 0 and 1. Therefore, the optimal value of ${a_{c,j}}$ must be infinitely close to 0 or 1 when maximizing the objective function. In this way, the first subproblem with integer variables transforms into a nonlinear programming (NLP) problem with continuous variables, which can be solved by known optimization algorithms such as interior point method. In this work, we adopt the toolbox \textit{fmincon()} in MATLAB optimization toolbox to implement the interior point method.
It is noteworthy that the value of $\lambda$ can affect the convergence of the iterative algorithm. In particular, $\lambda$ should be large enough to satisfy the constraints, while an excessively large $\lambda$ can weaken the objective function $R(a_{c,j})$ \cite{penalty}. In line with \cite{lamvalue10time}, we set $\lambda$ $10^3$ times larger than the objective function. 

From the obtained optimal ${{\boldsymbol{A}}_1},{{\boldsymbol{A}}_2},\cdots $ and ${{\boldsymbol{A}}_{N_c}}$, we can get the indexes of the $n_t$ activated LEDs, and the rest $\left(N_{\rm{T}}-n_t\right)$ LEDs are in sleep mode. Due to the variation of the activated LEDs, the cell formation should be updated. In each update, $n_t$ LEDs are reallocated into  $N_c$ cells, the process of which is similar to the steps in \textit{LED association} stated before. The detailed procedure of cell formation update is described as follows:

1) Construct a new association matrix ${\boldsymbol{M}} \in {\mathcal{R}^{{N_{\rm{R}}} \times {n_t}}}$, whose value is the channel gains between ${N_{\rm{R}}}$ users and ${n_t}$ LEDs. For each row $i$, find the strongest LOS channel gain at the $j_i$ column. Then, we associate user $i$ with LED $j_i$ since the corresponding channel gain is the strongest for user $i$, and set the $j_i$th column of ${\boldsymbol{M}}$ to ${\boldsymbol{0}}$. Finally, there are ${N_{\rm{R}}}$ ${\boldsymbol{0}}$ columns in ${\boldsymbol{M}}$, which means those ${N_{\rm{R}}}$ LEDs have been allocated to cells.

2) For the remaining non-zero $\left(n_t-{N_{c,\rm{R}}}\right)$ columns of ${\boldsymbol{M}}$, find the strongest channel gain amongst each column. Specifically, for the $m$th column, if the index of the strongest channel gain amongst $m$th column  of ${\boldsymbol{M}}$ is $n$, we associate the $m$th LED to the cell where the $n$th user is in, and set $m$th column to ${\boldsymbol{0}}$. Repeat this step until all the columns of ${\boldsymbol{M}}={\boldsymbol{0}}$.

Finally, all the $n_t$ activated LEDs are allocated into ${N_c}$ cells. The cell formation update is finished.
\vspace{-0.cm}
\subsection{Precoding Matrix Design}
\vspace{-0.cm}
Given the LED selection matrix ${\boldsymbol{A}}_c$ obtained via solving the first subproblem, the second subproblem can be written as
\vspace{-0.1cm}
\begin{align}
\mathop {\max }\limits_{{{\boldsymbol{W}}_c}} &\quad\quad\quad\quad\quad R, \label{sub2withW}\\
\rm{s.t.}&\quad\quad \eqref{maxsumratea},\eqref{maxsumrateb},\eqref{maxsumratee}.\notag
\end{align}
Define the channel matrix ${\widehat {\boldsymbol{H}}_c} = {{\boldsymbol{H}}_c}{{\boldsymbol{A}}_c} \in {\mathcal{R}^{{N_{c,\rm{R}}} \times {N_{c,\rm{T}}}}}$. From \eqref{maxsumratea}, we have ${\widehat {\boldsymbol{H}}_c}{{\boldsymbol{W}}_c} = {\rm{diag}}\left\{ {\sqrt{{\boldsymbol{q}}_c}} \right\}$, and thus the constraints \eqref{maxsumratea} and \eqref{maxsumrateb} can be combined as
\vspace{-0.cm}
\begin{equation}
{\left\| {{{\left[ {{\widehat {\boldsymbol{H}}_c^\dag } {\rm{diag}}\left\{ {\sqrt{{\boldsymbol{q}}_c}} \right\}} \right]}_{(j,:)}}} \right\|_1} \le \Delta I,\forall j \in \left\{ {1,2, \cdots ,{N_{\rm{c,T}}}} \right\}, \label{precodwithq}
\end{equation}
where ${\widehat {\boldsymbol{H}}_c^\dag }$ is the generalized inverse matrix of ${\widehat {\boldsymbol{H}}_c}$.

The constraint \eqref{precodwithq} is non-convex since $\sqrt {{q_{c,i}}}$ is concave. To make \eqref{precodwithq} convex, we first square it as follows
\begin{equation}
{\left\| {{{\left[ {{\widehat {\boldsymbol{H}}_c^\dag}{\rm{diag}}\left\{ {\sqrt {{{\boldsymbol{q}}_c}} } \right\}} \right]}_{(j,:)}}} \right\|_{\rm{1}}}^{\rm{2}} \le \Delta {I^2}. \label{squQ1}
\end{equation}
Then according to the mean inequality\cite{ZFdelMUI}, \eqref{squQ1} obeys
\begin{footnotesize}
\begin{equation}
\frac{{{{\left\| {{{\left[ {\widehat {\boldsymbol{H}}_c^\dag {\rm{diag}}\left\{ {\sqrt {{{\boldsymbol{q}}_c}} } \right\}} \right]}_{j,:}}} \right\|}_{\rm{1}}}^{\rm{2}}}}{{{N_{c,{\rm{R}}}}}} \le {\left\| {{{\left[ {\widehat {\boldsymbol{H}}_c^\dag {\rm{diag}}\left\{ {{{\left( {\sqrt {{{\boldsymbol{q}}_c}} } \right)}^2}} \right\}\widehat {\boldsymbol{H}}{{_c^\dag }^{\rm{T}}}} \right]}_{(j,:)}}} \right\|_{\rm{1}}}
.\label{meaninq}
\end{equation}
\end{footnotesize}Combining \eqref{squQ1} and \eqref{meaninq}, \eqref{precodwithq} can be replaced by a stronger inequality written as
\begin{small}
\begin{equation}
\hspace{-0.2cm}\begin{array}{*{20}{c}}
{{{\left\| {{{\left[ {\widehat {\boldsymbol{H}}_c^\dag {\rm{diag}}\left\{ {{{\boldsymbol{q}}_c}} \right\}\widehat {\boldsymbol{H}}{{_c^\dag }^{\rm{T}}}} \right]}_{(j,:)}}} \right\|}_1} \le \frac{{\Delta {I^2}}}{{{N_{c,\rm{R}}}}}}&{\forall j \in \left\{ {1,2, \cdots ,{N_{c,\rm{T}}}} \right\}.}
\end{array} \label{strongQ}
\end{equation}
\end{small}
\begin{algorithm}[t]
	\caption{Subgradient Method to Solve the Dual Problem of \eqref{sub2withq}}
	\vspace{-0.cm}
	\begin{algorithmic}
		\REQUIRE
		Iteration counter $t=0$, step size parameter $a>0$, dual variables $\mu _j^{\left( 0 \right)} > 0$, $\lambda _i^{\left( 0 \right)} > 0$.
		\REPEAT
		\STATE Set step size ${\theta ^{\left( t \right)}} = \frac{a}{{\sqrt t }}$;
		\FOR{$i=1$ to $N_{\rm{c,R}}$}
		\STATE Calculate $q_{c,i}^{*\left( t \right)} = f\left( {\mu _j^{\left( t \right)},\lambda _i^{\left( t \right)}} \right)$ by \eqref{optimalq};	
		\STATE Update ${\lambda _i^{\left( t+1 \right)}}$ by \eqref{gradlam};
		\ENDFOR
		\FOR{$i=1$ to $N_{\rm{c,T}}$} 	
		\STATE Update ${\mu _j^{\left( t+1 \right)}}$ by \eqref{gradmu};
		\ENDFOR	
		\STATE Update $t = t+1$.
		\UNTIL ${\left| {{R^{\left( t \right)}} - {R^{\left( {t - 1} \right)}}} \right|^2} \le {{\varepsilon}_2}$ or $t=T$, where ${{\varepsilon}_2}$ and $T$ are a predefined threshold of accuracy and a predefined maximum number of iterations, respectively.
	\end{algorithmic}
\end{algorithm}Therefore, the second subproblem with respect to ${{\boldsymbol{q}}_c}$ can be rewritten as
\vspace{-0.cm}
\begin{align}
&\hspace{-0.3cm}\mathop {\max }\limits_{{{\boldsymbol{q}}_c}} R\left({\boldsymbol{q}}_c\right) = \frac{1}{2}\sum\limits_{c = 1}^{{N_c}} {\sum\limits_{i = 1}^{{N_{c,\rm{R}}}} {\log } } \left( {1 + \frac{{2{{\left( {\gamma \zeta } \right)}^{\rm{2}}}}}{{\pi e\left( {{\delta _{c,i}} + \sigma _{c,i}^2} \right)}}{q_{c,i}}} \right),\label{sub2withq}
\\&\rm{s.t.}\quad\quad\quad\quad\quad\quad\quad\quad\eqref{maxsumratee},\eqref{strongQ},\notag
\end{align}
where \begin{footnotesize}${\delta _{c,i}} = \frac{{{{\left( {\gamma \varsigma } \right)}^2}}}{3}\sum\limits_{c' \ne c}^{{N_c}} {\sum\limits_{j \in {{\mathcal{U}}_{c'}}}^{{N_{c',\rm{R}}}} {{{\left( {{{\boldsymbol{H}}_{c',(i,:)}}{{\boldsymbol{A}}_{c'}}{{\left[ {\widehat {\boldsymbol{H}}_{c'}^\dag \rm{diag}\left\{ {\sqrt {{{\boldsymbol{q}}_{c'}}} } \right\}} \right]}_{(:,j)}}} \right)}^2}} }$\end{footnotesize}. 
Since the Hessian of $R\left({\boldsymbol{q}}_c\right)$ is a negative definite matrix, the objective function of \eqref{sub2withq} is concave \cite{boydconvex}[3.1.4]. Meanwhile, the constraint \eqref{maxsumratee} is obviously a linear function of ${q_{c,i}}$. In addition, in the constraint \eqref{strongQ}, $g\left( {{{\boldsymbol{q}}_c}} \right)={{\left\| {{{\left[ {\widehat {\boldsymbol{H}}_c^\dag {\rm{diag}}\left\{ {{{\boldsymbol{q}}_c}} \right\}\widehat {\boldsymbol{H}}{{_c^\dag }^{\rm{T}}}} \right]}_{(j,:)}}} \right\|}_1}$ is the $L_1$ norm of the function $f\left( {{{\boldsymbol{q}}_c}} \right) = {\left[ {\widehat {\bf{H}}_c^\dag {\rm{diag}}\left\{ {{{\boldsymbol{q}}_c}} \right\}\widehat {\bf{H}}{{_c^\dag }^{\rm{T}}}} \right]_{\left( {j,:} \right)}}$. Since $g\left( {{{\boldsymbol{q}}_c}} \right)$ is convex with respect to $f\left( {{{\boldsymbol{q}}_c}} \right)$ \cite{boydconvex}[3.1.5], and $f\left( {{{\boldsymbol{q}}_c}} \right)$ is a linear function of ${{\boldsymbol{q}}_c}$, the constraint \eqref{strongQ} is convex with respect to ${{\boldsymbol{q}}_c}$. Therefore, \eqref{sub2withq} is a convex problem\cite{boydconvex}[4.2.1]. Although optimization software such as CVX provides efficient tools to solve convex problems, it may not be able to solve problems with complex structure and a large number of variables\cite{cvxcannot}. For example, a complex structured convex problem may violate the disciplined convex programming ruleset required by CVX\cite{cvxusrguide}. Therefore, we use the Lagrangian dual method to solve \eqref{sub2withq}, and its Lagrangian is written as
\vspace{-0.cm}
\begin{equation}
\hspace{-0.4cm}\begin{array}{l}
L\left( {\boldsymbol{q}},{\boldsymbol{\mu}},{\boldsymbol{\lambda}} \right) = \frac{1}{2}\sum\limits_{c = 1}^{{N_c}} {\sum\limits_{i = 1}^{{N_{c,\rm{R}}}} {\log \left( {1 + \frac{{2{{\left( {\gamma \zeta } \right)}^{\rm{2}}}}}{{\pi e\left( {{\delta _{c,i}} + \sigma _{c,i}^2} \right)}}{q_{c,i}}} \right)} } \\
- \sum\limits_{j = 1}^{{N_{c,\rm{T}}}} {{\mu _j}} \left[ {h_{j,i}^\dag \left( {\sum\limits_{m = 1}^{{N_{c,\rm{R}}}} {h{{_{i,m}^\dag }^{\rm{T}}}} } \right){q_{c,i}} - \frac{{\Delta {I^2}}}{{{N_{c,\rm{R}}}}}} \right] + \sum\limits_{i = 1}^{{N_{c,\rm{R}}}} {{\lambda _i}{q_{c,i}}},
\end{array} \label{lagfunc}
\end{equation} where ${\boldsymbol{q}}=\left\{ {{q_{c,i}}} \right\},{\boldsymbol{\mu}}  = \left\{ {{\mu _j}} \right\},{\boldsymbol{\lambda}}  = \left\{ {{\lambda _i}} \right\}, \forall c \in \left\{ {1, \cdots ,{N_c}} \right\},\forall i = \left\{ {1,2, \cdots ,{N_{c,\rm{R}}}} \right\}, \forall j = \left\{ {1,2, \cdots ,{N_{c,\rm{T}}}} \right\}$, and ${{\mu _j}}$ and ${{\lambda _i}}$ are dual variables for constraint \eqref{strongQ} and \eqref{maxsumratee}, respectively. From Karush-Kuhn-Tucker (KKT) conditions

\begin{small}
\vspace{-0.8cm}
\begin{align} \label{kkt}
\hspace{-0.cm}
\begin{split}
&\frac{{{m_{c,i}}}}{{2\ln 2\left( {1 + {m_{c,i}}{q_{c,i}}} \right)}}
- \sum\limits_{j = 1}^{{N_{c,\rm{T}}}} {{\mu _j}\left( {h_{c,\left( {j,i} \right)}^\dag \sum\limits_{m = 1}^{{N_{c,\rm{R}}}} {h_{c,\left( {i,m} \right)}^{\dag {\rm{T}}}} } \right)}\\
& + {{\lambda _i}} = 0, \qquad\qquad\qquad\qquad\qquad\qquad\qquad \forall i = \left\{ {1,2, \cdots ,{N_{c,\rm{R}}}} \right\},\\
&{{\mu _j}\left( {h_{c,\left({j,i}\right)}^\dag \sum\limits_{i = 1}^{{N_{c,\rm{R}}}}{h_{c,\left( {i,m} \right)}^{\dag {\rm{T}}}{q_{c,i}} - \frac{{\Delta {I^2}}}{{{N_{c,\rm{R}}}}}} } \right) = 0},\forall j = \left\{ {1,2, \cdots ,{N_{c,\rm{T}}}} \right\},\\
&{\mu _j} \ge 0,\quad\quad\quad\qquad\qquad\qquad\qquad\qquad\qquad \forall j = \left\{ {1,2, \cdots ,{N_{c,\rm{T}}}} \right\},\\
&{\lambda _j}{q_{c,i}}{\rm{ = }}0,\qquad\qquad\qquad\qquad\qquad\qquad\qquad\forall i = \left\{ {1,2, \cdots ,{N_{c,\rm{R}}}} \right\},\\
&{\lambda _j} \ge 0,\quad\qquad\qquad\qquad\qquad\qquad\qquad\qquad\forall i = \left\{ {1,2, \cdots ,{N_{c,\rm{R}}}} \right\},
\end{split}
\end{align}
\end{small}where ${m_{c,i}} = \frac{{2{{\left( {\gamma \zeta } \right)}^{\rm{2}}}}}{{\pi e\left( {{\delta _{c,i}} + \sigma _{c,i}^2} \right)}}$. By solving \eqref{kkt}, the optimal $q_{c,i}^*$ can be obtained as
\begin{small}
\begin{align}
\hspace{-0.2cm} q_{c,i}^* &= f\left( {{\mu _j},{\lambda _i}} \right) \notag\\
&=
\frac{1}{{2\ln 2\left( {\sum\limits_{j = 1}^{{N_{c,\rm{T}}}} {{\mu _j}\left( {h_{c,\left( {j,i} \right)}^\dag \sum\limits_{m = 1}^{{N_{c,\rm{R}}}} {h_{c,\left( {i,m} \right)}^{\dag {\rm{T}}}} } \right) -  {{\lambda _i}} } } \right)}} - \frac{1}{{{m_{c,i}}}}. \label{optimalq}
\end{align}
\end{small}Plugging \eqref{optimalq} into \eqref{lagfunc}, the Lagrangian dual problem is
\begin{alignat}{2}
&\mathop {\min }\limits_{{\mu _j},{\lambda _i}} \quad &&L({\mu _j},{\lambda _i}),\label{lgdual}\\
&\rm{s.t.}&&\begin{array}{*{20}{c}}
{{\mu _j} \ge 0},&{\forall j \in \left\{ {1,2, \cdots ,{N_{c,\rm{T}}}} \right\},}
\end{array}\tag{\ref{lgdual}{a}}\label{lgduala}\\
&&&\begin{array}{*{20}{c}}
{{\lambda _i} \ge 0},&{\forall i \in \left\{ {1,2, \cdots ,{N_{c,\rm{R}}}} \right\}.}
\end{array}\tag{\ref{lgdual}{b}}\label{lgdualb}
\end{alignat}
\begin{algorithm}[t]
	\caption{Iterative Algorithm to Solve \eqref{maxsumrate}}
	\begin{algorithmic}[1]
		\REQUIRE
		\STATE Iteration counter $t=0$, channel matrices ${\boldsymbol{H}}_c$ and ${\sigma _{c,i}^2}$. Initialize precoding matrices ${\boldsymbol{W}}_c^{(0)}{\boldsymbol{W}}_c^{(0){\rm{T}}}$, e.g., ${\boldsymbol{W}}_c^{(0)}{\boldsymbol{W}}_c^{(0){\rm{T}}} = {\boldsymbol{I}}$.
		\STATE Initialize the UC cell formation. 		
		\REPEAT
		\STATE Update the suboptimal LED selection ${\boldsymbol{A}}_c^{(t+1)}$ with fixed ${\boldsymbol{W}}_c^{(t)}$ by solving \eqref{provequal} using \textit{fmincon()} in MATLAB optimization toolbox;
		\STATE Reallocate the activated $n_t$ LEDs into $N_c$ cells and update the cell formation with the obtained ${\boldsymbol{A}}_c^{(t+1)}$;
		\STATE Solve \eqref{sub2withq} to obtain the optimal precoding matrix ${\boldsymbol{W}}_c^{(t+1)}$ with the obtained ${\boldsymbol{A}}_c^{(t+1)}$ by Lagrangian dual method;
		\STATE Calculate ${R^{\left( t+1 \right)}}$ by \eqref{capa2}. If ${R^{(t + 1)}} < {R^{(t)}}$, set ${R^{(t + 1)}} = {R^{(t)}}$ ;
		\STATE Update $t = t+1$.
		\UNTIL ${\left| {{R^{\left( t+1 \right)}} - {R^{\left( t \right)}}} \right|^2} \le {{\varepsilon}_3}$, where ${{\varepsilon}_3}$ and $t=T$ are a predefined  and a predefined maximum number of iterations.
	\end{algorithmic}
\end{algorithm}The dual variables ${\mu _j}$ and ${\lambda _i}$ can be easily solved by the subgradient descent method in an iterative manner as shown in Algorithm 1. In the $(t+1)$th iteration, the values of ${\mu _j}$ and ${\lambda _i}$ are updated according to

\begin{small}
\begin{equation}
\mu _j^{\left( {t + 1} \right)} = {\left[ {\mu _j^t - {\theta ^t}\left( {h_{c,\left( {j,i} \right)}^\dag \sum\limits_{i = 1}^{{N_{c,\rm{R}}}} {h_{c,\left( {i,m} \right)}^{\dag {\rm{T}}}f\left( {{\mu _j},{\lambda _i}} \right) - \frac{{\Delta {I^2}}}{{{N_{c,\rm{R}}}}}} } \right)} \right]^ + ,} \label{gradmu}
\end{equation}
\end{small}
and
\begin{equation}
\lambda _i^{(t + 1)} = {\left[ {\lambda _i^t - {\theta ^t}f\left( {\mu _j,\lambda _i} \right)} \right]^{\rm{ + }}}, \label{gradlam}
\end{equation}
 where ${\left[ {\rm{*}} \right]^{\rm{ + }}}{\rm{ = max(*,0)}}$ and ${\theta ^{\left( t \right)}} = \frac{a}{{\sqrt t }}$ is a dynamic stepsize with $a$ being a constant which must be sufficiently small to ensure that the algorithm converges to an optimal value\cite{suffismallSS}. This iterative process stops when ${\left| {{R^{\left( t \right)}} - {R^{\left( {t - 1} \right)}}} \right|^2} \le {{\varepsilon}_2}$, where ${{\varepsilon}_2}$ is a predefined threshold of accuracy. After obtaining the optimal ${\boldsymbol{q}}_c$, the optimal ${\boldsymbol{W}}_c$ is calculated by \eqref{eq10}.

With regards to the convergence, we first prove the convergence of Algorithm 1 and then introduce the convergence of Algorithm 2. In particular, the convergence of Algorithm 1 is shown in the following proposition.

\noindent
\textbf{Proposition 2}: Algorithm 1 always converges.

\textbf{Proof}: See Appendix A.\hfill $\blacksquare$

Then we substitute the obtained ${\boldsymbol{W}}_c$ into the first subproblem iteratively until a convergent solution of \eqref{maxsumrate} is found. The iterative algorithm is summarized in Algorithm 2, which is guaranteed to converge. See Appendix B for detailed proof.

\subsection{Complexity Analysis}

The complexity analysis of the proposed iterative algorithm is divided into two parts according to the two subproblems. The first subproblem includes UC cell formation and LED selection. For UC cell formation, each user is assigned to an anchored LED to form distance-based clusters in the first step. Hence, the complexity of this step is $\mathcal{O}\left({N_{\rm{R}}}\right)$. Then, the remaining ${N_{\rm{T}}}-{N_{\rm{R}}}$ LEDs are associated to the clusters formed in the first step with complexity of $\mathcal{O}\left({N_{\rm{T}}}-{N_{\rm{R}}}\right)$. Therefore, the total computational complexity of the UC cell formation is $\mathcal{O}\left({N_{\rm{T}}}\right)$. After the cell formation, the LED selection problem is solved by \textit{fmincon()} function in MATLAB toolbox, which implements the interior point method. The computational complexity of the interior point method is related to the total number of constraints $\left( {{N_{\rm{T}}} + 1} \right)$, the initial point ${t^0}$, the accuracy ${\varepsilon}_1$ and the gradient of step size $\xi $, written as $\mathcal{O}\left(\frac{{{{\log }_2}\left( {{{\left( {{N_{\rm{T}}} + 1} \right)} \mathord{\left/
					{\vphantom {{\left( {{N_{\rm{T}}} + 1} \right)} {{t^0}{\varepsilon _1}}}} \right.
					\kern-\nulldelimiterspace} {{t^0}{\varepsilon _1}}}} \right)}}{{{{\log }_2}\xi }}\right)$ \cite{boydconvex}.  Hence, the computational complexity of solving the first subproblem is  $\mathcal{O}\left(\frac{{{{\log }_2}\left( {{{\left( {{N_{\rm{T}}} + 1} \right)} \mathord{\left/
					{\vphantom {{\left( {{N_{\rm{T}}} + 1} \right)} {{t^0}{\varepsilon _1}}}} \right.
					\kern-\nulldelimiterspace} {{t^0}{\varepsilon _1}}}} \right)}}{{{{\log }_2}\xi }} + {N_{\rm{T}}}\right) $.
		
For the second subproblem, the Lagrange dual problem is solved by a subgradient descent method, whose complexity in terms of accuracy ${{\varepsilon}_2}$ is $\mathcal{O}\left( {{\rm{1}} \mathord{\left/
		{\vphantom {{\rm{1}} {\sqrt \varepsilon  }}} \right.
		\kern-\nulldelimiterspace} {\sqrt {{\varepsilon}_2}}} \right)$. Each iteration requires the following computational operation: (i) computing \begin{footnotesize}${\delta _{c,i}} = \frac{{{{\left( {\gamma \varsigma } \right)}^2}}}{3}\sum\limits_{c' \ne c}^{{N_c}} {\sum\limits_{j \in {{\mathcal{U}}_{c'}}}^{{N_{c',\rm{R}}}} {{{\left( {{{\boldsymbol{H}}_{c',(i,:)}}{{\boldsymbol{A}}_{c'}}{{\left[ {{\boldsymbol{H}}_{c',sel}^\dag \rm{diag}\left\{ {\sqrt {{{\bf{q}}_{c'}}} } \right\}} \right]}_{:,j}}} \right)}^2}} }$\end{footnotesize} with $\mathcal{O}\left({N_{\rm{T}}^2 + {N_{\rm{T}}}{N_{\rm{R}}} + {N_{\rm{T}}}N_{\rm{R}}^2}\right)$\cite{CompuComplx}; (ii) computing $\sigma _{c,i}^2$ includes $ {{{\boldsymbol{H}}_{c,(i,:)}}{\boldsymbol{I}}_{\rm{B}}^c}$ with $\mathcal{O}\left({N_{\rm{T}}}\right)$; (iii) computing the pseudo-inverse ${\boldsymbol{H}}_c^\dag$ by SVD decomposition with $\mathcal{O}\left( {{N_{\rm{T}}}N_{\rm{R}}^2} \right)$\cite{SVD}. Hence the computational complexity of the second subproblem is $\mathcal{O}\left({{2{N_{\rm{T}}}N_{\rm{R}}^2} \mathord{\left/
		{\vphantom {{2{N_{\rm{T}}}N_{\rm{R}}^2} {\sqrt {{\varepsilon}_2} }}} \right.
		\kern-\nulldelimiterspace} {\sqrt {{\varepsilon}_2} }}\right)$.

 Define the number of iterations that Algorithm 2 needs to converge as $L$, the total computational complexity is $\mathcal{O}\left( L\left( {{{\frac{{{{\log }_2}\left( {{{\left( {{N_{\rm{T}}} + 1} \right)} \mathord{\left/
 								{\vphantom {{\left( {{N_{\rm{T}}} + 1} \right)} {{t^0}{\varepsilon _1}}}} \right.
 								\kern-\nulldelimiterspace} {{t^0}{\varepsilon _1}}}} \right)}}{{{{\log }_2}\xi }} + 2{N_{\rm{T}}}N_{\rm{R}}^2} \mathord{\left/
 			{\vphantom {{{{\log }_2}\left( {{1 \mathord{\left/
 									{\vphantom {1 \varsigma }} \right.
 									\kern-\nulldelimiterspace} \varsigma }} \right) + 2{N_{\rm{T}}}N_{\rm{R}}^2} {\sqrt {{\varepsilon}_2}  }}} \right.
 			\kern-\nulldelimiterspace} {\sqrt {{\varepsilon}_2} }}} + {N_{\rm{T}}} \right) \right)$. In comparison, if the first subproblem is solved by exhaustive search, the computational complexity is $\mathcal{O}\left(L\left( {{{{2^{{N_{\rm{T}}}}} + 2{N_{\rm{T}}}N_{\rm{R}}^2} \mathord{\left/
 			{\vphantom {{{2^{{N_{\rm{T}}}}} + 2{N_{\rm{T}}}N_{\rm{R}}^2} {\sqrt {{\varepsilon}_2}  }}} \right.
 			\kern-\nulldelimiterspace} {\sqrt {{\varepsilon}_2} }}} + {N_{\rm{T}}} \right) \right)$, which is significantly higher than our proposed algorithm.

\begin{figure}[t]
	\vspace{-0.cm}
	\centerline{\includegraphics[width=1\linewidth]{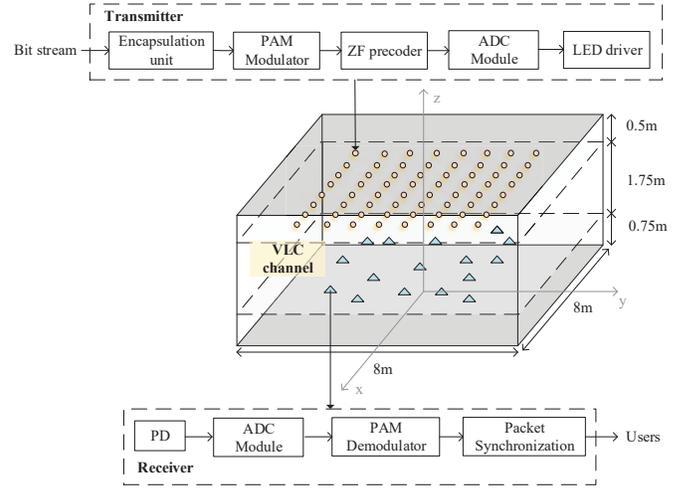}}
	\caption{Implementaion of TASP-HD.}
	\label{imple}
	\vspace{-0.cm}
\end{figure}

\subsection{Implementation of TASP-HD}

The implementation of TASP-HD is presented in Fig \ref{imple}. At the transmitter, multiple LEDs provide both communication and illumination functions. Each transmitter consists of an encapsulation unit, a modulator, a precoder and an LED driver. An encapsulation unit is used to create data packet that includes the start frame
delimiter (SFD) and the identity information of the transmitter. The SFD consists of a leading
bit and the synchronization code. Then, the data packet is modulated with modulation scheme by a microcontroller. In this paper, PAM is adopted. After modulation, the PAM symbol passes through a precoder. ZF precoding is adopted to eliminate intra-cell interference, and a DC bias is added to ensure the signal is positive and real.

At the receiver, the device having a front PD such as smartphones and panel computers can be used. The PD receives the visible light from each LED via LOS channel in an individual time slot and converts the incident photon into an electron/electric current. Then, the sampling of the analog signal is performed using the analog-to-digital converter (ADC) module to obtain the transmitted bits. After the demodulation and packet synchronization modules, the data is transmitted to users. This completes the implementation of the proposed TASP-HD.
\section{Simulation Results}
\begin{table}[t]
	\vspace{0cm}
	\small
	\caption{Simulation Parameters.}
	\vspace{-0.cm}
	\begin{center}
		\setlength{\belowcaptionskip}{0pt}
		\begin{tabular}{lp{66pt}<{\centering}}
			\hline		
			\multicolumn{2}{c}{ Environment-related Parameters}\\
			\hline
			\textbf{Parameter}& \textbf{Value}\\
			\hline
			Room size  & 8 m$\times$8 m$\times$3 m\\		
			{Number of LEDs, $N_{\rm{T}}$} & 36, 64\\
			Number of users, $N_{\rm{R}}$ & 12, 16\\
			Semiangle of half power, ${\Phi _{{1 \mathord{\left/
							{\vphantom {1 2}} \right.
							\kern-\nulldelimiterspace} 2}}}$ & ${80^ \circ }$\\
			Detect area of PDs, ${A_r}$ & 1 $\rm{c{m^2}}$ \\
			FOV of the PD, $\Psi $ & ${60^ \circ }$\\
			Optical filter gain, ${{T_s}({\psi _{c,j}})}$ & 1\\
			Refractive index of optical concentrator, $\kappa$ & 1\\
			Height from TXs to RX, $h$ & 1.75 m\\
			Dynamic range of current, $[{I_{\rm{l}}},{I_{\rm{h}}}]$ & [0 A, 2 A]\\
			{Illumination range of ISO standard}& [300 lx, 1500 lx]\\
			System bandwidth, $B$ & 100 MHz\\
			PD responsivity, $\gamma $ & 0.54 A/W\\
			Electrical-to-optical conversion coefficient, $\zeta$ & 0.44 W/A\\
			Ambient light photocurrent, ${\chi _{amb}}$ & 10.93 A/$\rm{m}^2\cdot$Sr \\
			Preamplifier noise current density, ${i_{amb}}$ & 5 pA/${\rm{H}}{{\rm{z}}^{ - 1/2}}$\\
			\hline
			\multicolumn{2}{c}{ Baseline Algorithm-related Parameters}\\
			\hline
			\textbf{Parameter} & \textbf{Value}\\
			{Number of CV(RMSE) sample points, $K$ }& 729\\
			{Penalty parameter in \eqref{provequal}, $\lambda$}& $10^5$\\
			{Initial stepsize of subgradient method, $a$ }& $0.01$\\
			{Convergence threshold, ${{\varepsilon}_2},{{\varepsilon}_3}$ }& $0.001$\\
			\hline
		\end{tabular}
	\end{center}\label{tab2}
	\vspace{-0.cm}
\end{table}
In this section, a MC MU-MISO VLC system employing TASP-HD in a 8 m$\times$8 m$\times$3 m square room is considered, where ${N_{\rm{R}}}$ users are randomly distributed on a plane 0.75 m above the floor, and ${N_{\rm{T}}}$ LEDs are evenly installed on a plane at 2.5 m height. In this section, when ${N_{\rm{T}}}=36$, ${N_{\rm{R}}}$ is set to 12; and when ${N_{\rm{T}}}=64$, ${N_{\rm{R}}}$ is set to 16. The minimum and maximum allowed currents are set to 0 A and 2 A. In addition, the conventional DD\cite{dd2} and AD\cite{ad2} are adopted as baselines to compare with the TASP-HD. The detailed simulation parameters are listed in \textbf{Table II}.
\begin{figure}[t]
	\subfigure[\small $N_{\rm{T}}=36$]{
		\begin{minipage}{1\linewidth}
			\hspace{-0.cm}
			\centerline{\includegraphics[width=0.8\linewidth]{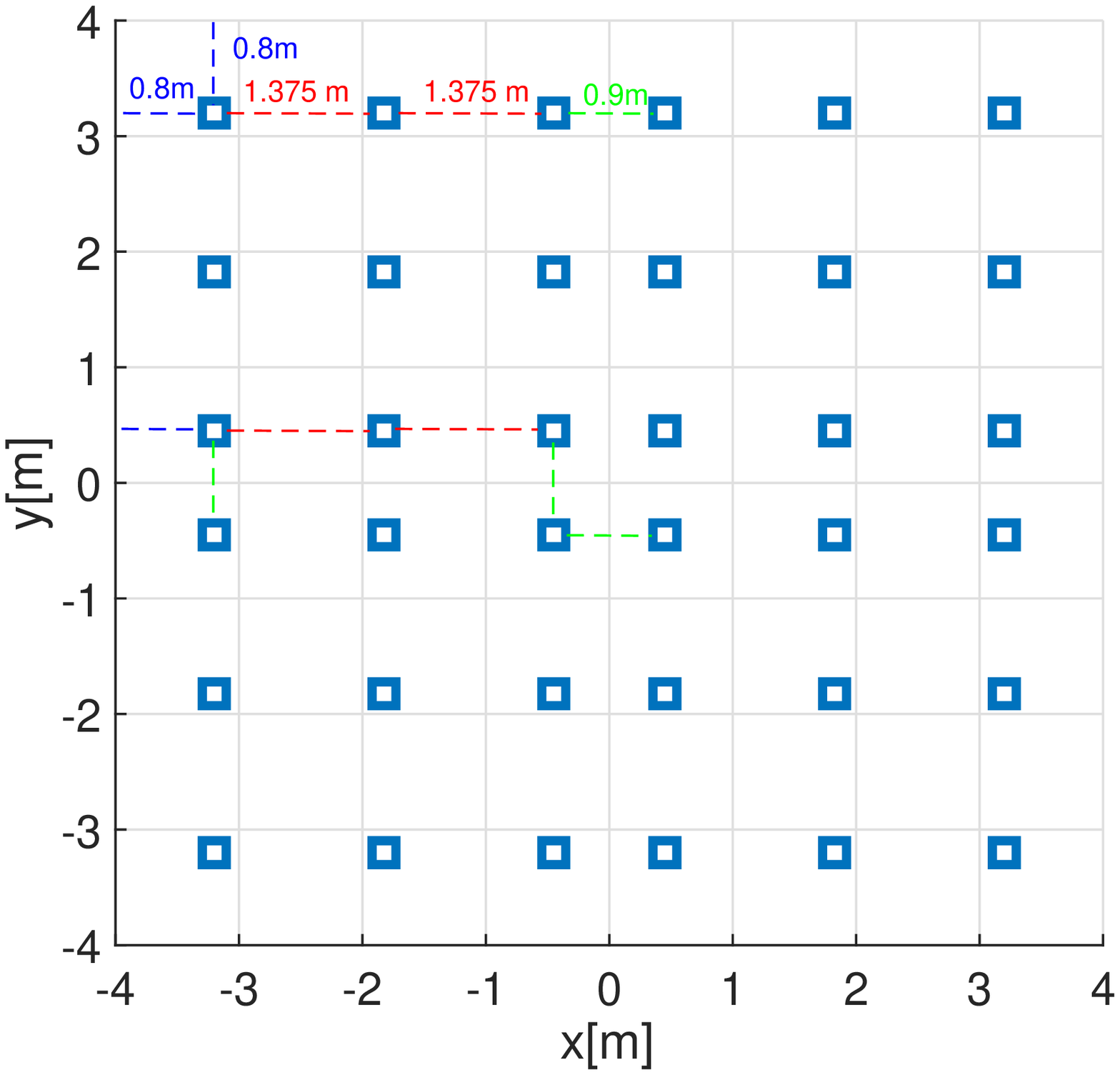}}\\
			\vspace{-0.5cm}
		\end{minipage}%
	}%
	\vspace{-0.cm}
	\subfigure[\small $N_{\rm{T}}=64$]{
		\begin{minipage}{1\linewidth}
			\vspace{-0.cm}
			\centerline{\includegraphics[width=0.8\linewidth]{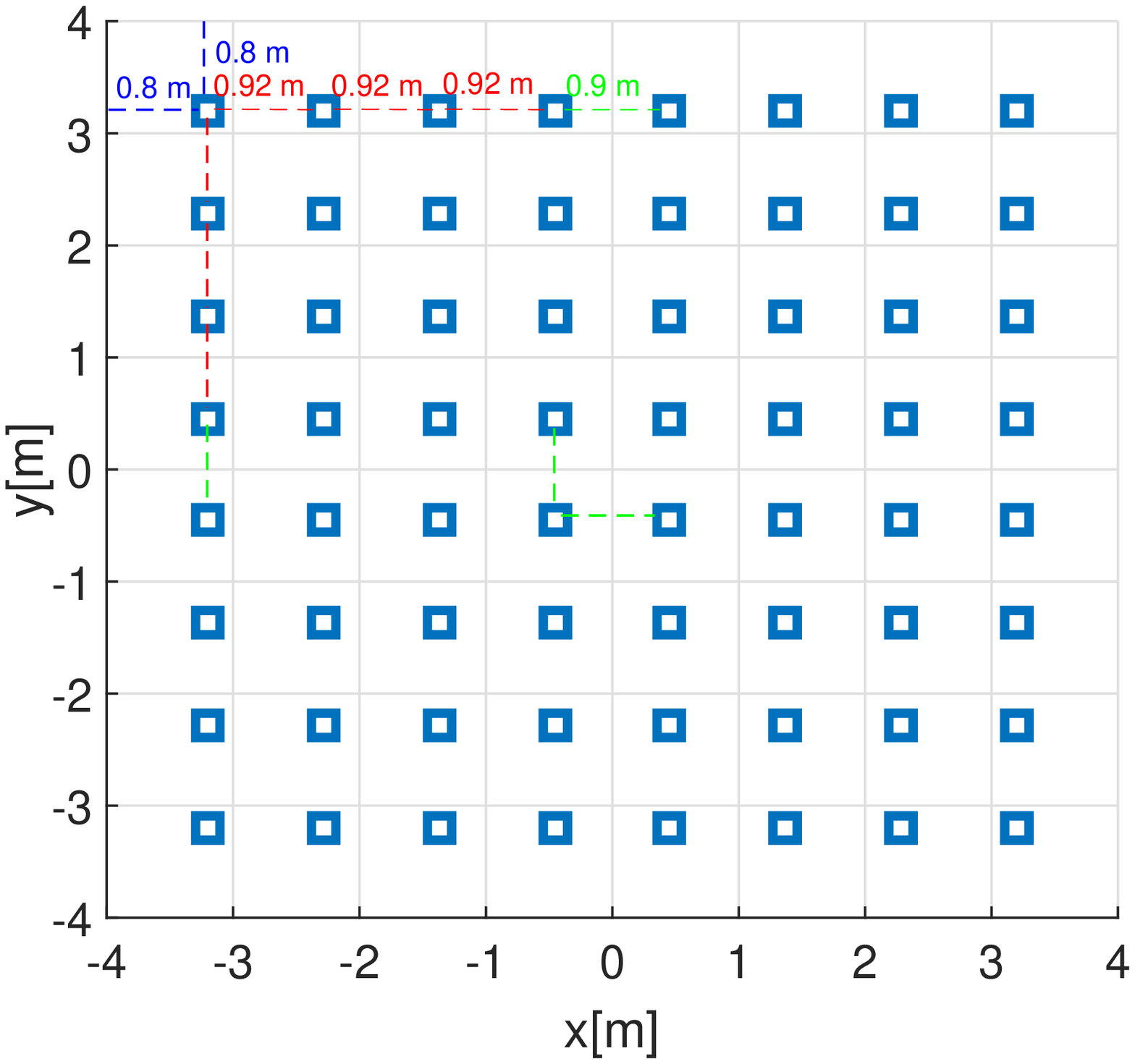}}
			\vspace{-0.3cm}
		\end{minipage}%
	}%
	\vspace{-0.3cm}
	\caption{LED distribution when $N_{\rm{T}}=36,64$.}\label{figLEDloc}
	\vspace{-0.cm}
\end{figure}
\begin{figure}[t]
	\vspace{-0.cm}
	\centerline{\includegraphics[width=0.8\linewidth]{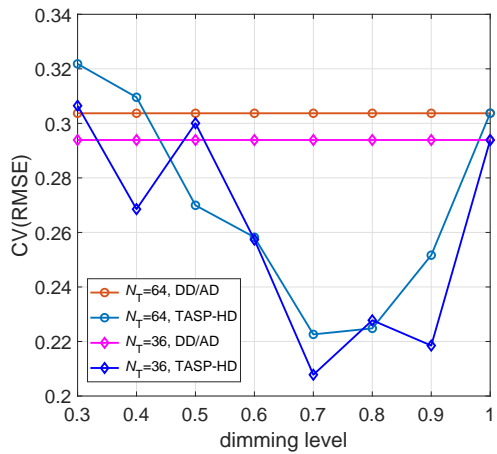}}
	\caption{Illumination uniformity of TASP-HD and baseline schemes under $\eta=30\%,50\%,70\%,90\%$ and $N_{\rm{T}}=36,64$.}
	\label{figCVrmse}
	\vspace{-0.cm}
\end{figure}
\begin{figure*}
	\vspace{-0.cm}
	\centerline{\includegraphics[width=1\linewidth]{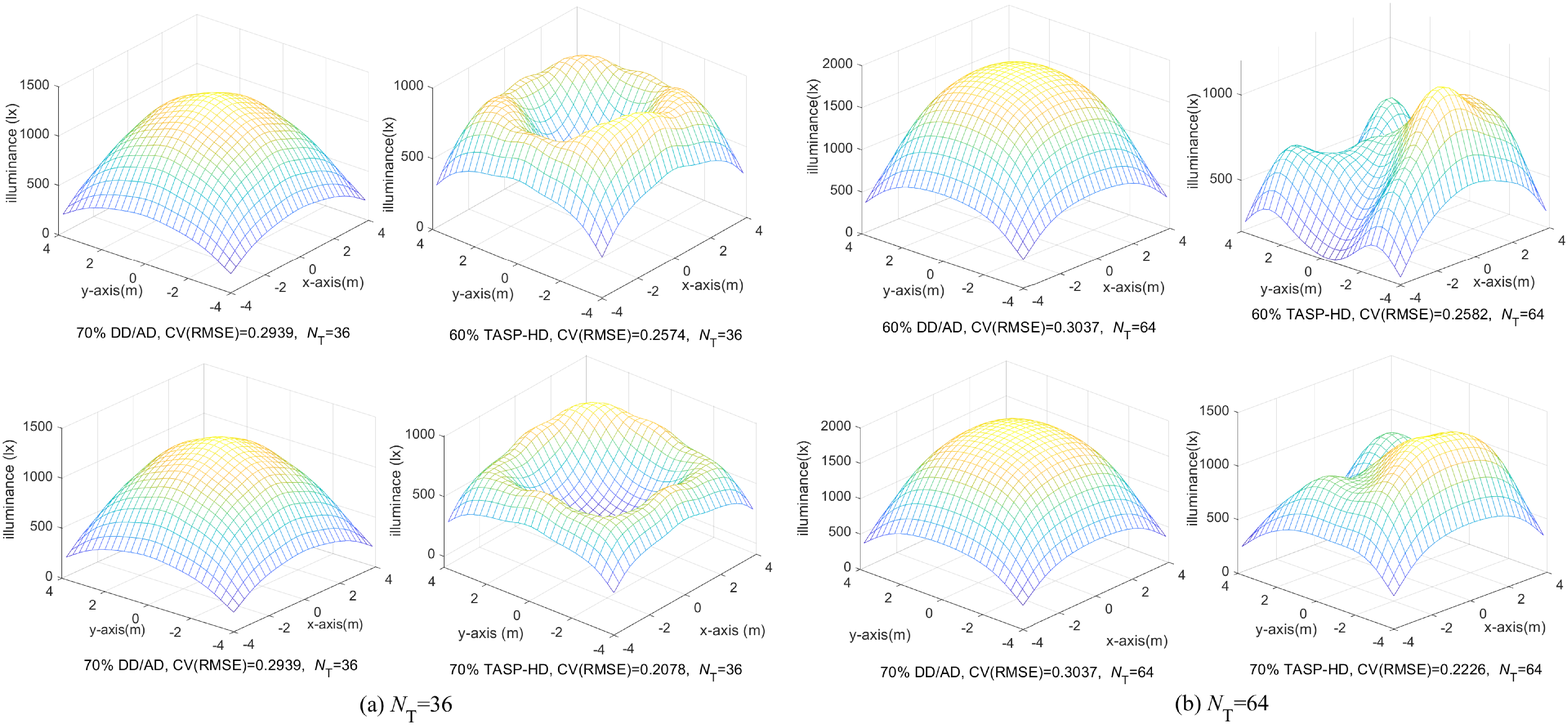}}
	\vspace{-0.cm}
	\caption{Illumination distribution of TASP-HD and baseline schemes under $\eta=60\%,70\%$ and $N_{\rm{T}}=36,64$.}
	\vspace{-0.cm}
	\label{fig3Duir}
\end{figure*}

\subsection{Illumination Performance Evaluations}
As shown in Fig. \ref{figLEDloc}, we consider two scenarios where $N_{\rm{T}}=36$ and $64$ LEDs are uniformly distributed in the square room. From Fig. \ref{figLEDloc} we can see that the predefined LED distribution with $N_{\rm{T}}=36$ is set to be more scattered and less intensive than that of $N_{\rm{T}}=64$. On the receiver plane, $K$ equally spaced sample points are considered, and the distance is set to 0.3 m. Therefore, $K={\left\lceil {\frac{{8{\rm{m}}}}{{0.3{\rm{m}}}}} \right\rceil ^2} = 729$. The illumination uniformity performance quantified by CV(RMSE) is illustrated in Fig. \ref{figCVrmse}. Since DD and AD activate all of the LEDs under different dimming levels, the illumination uniformity performance of DD and AD are identical. As shown in Fig. \ref{figCVrmse}, when $N_{\rm{T}}=36$ and $64$, the CV(RMSE) resulting from AD/DD are 0.2939 and 0.3037, respectively. This is because more LED can enhance the illuminance at the center of the room while the illuminance at the edge of the room remains the same, and this deteriorates the illumination uniformity performance. We can also observe that although the proposed TASP-HD has inferior illumination uniformity performance than DD/AD when dimming level is 30\%, it outperforms DD/AD in most cases when the dimming level increases, especially when the dimming level ranges from 60\% to 90\%. This is because at the lower dimming levels, the number of activated LEDs is small. Hence it is challenging for TASP-HD to achieve uniform illumination performance. When the dimming level increases, TASP-HD can activate more LEDs thus providing TASP-HD more possible configurations to achieve uniform illumination. In particular, when $N_{\rm{T}}=36$, TASP-HD provides higher illumination uniformity than DD/AD under 60\%-90\% dimming levels. Meanwhile, the values of CV(RMSE) of TASP-HD are 0.0365, 0.0861, and 0.0661 lower than that of DD/AD under 60\%, 70\%, and 80\% dimming levels, respectively. When $N_{\rm{T}}=64$, TASP-HD also achieves better illumination uniformity, whose CV(RMSE) values are 0.0455, 0.0811, and 0.0789 lower than DD/AD under 60\%, 70\%, and 80\% dimming levels, respectively. In addition, all of these three schemes attain the same value of CV(RMSE) under 100\% dimming level since they all activate all of the LEDs.
\vspace{-0.cm}
\begin{figure*}
	\vspace{0cm}
	\centerline{\includegraphics[width=1\linewidth]{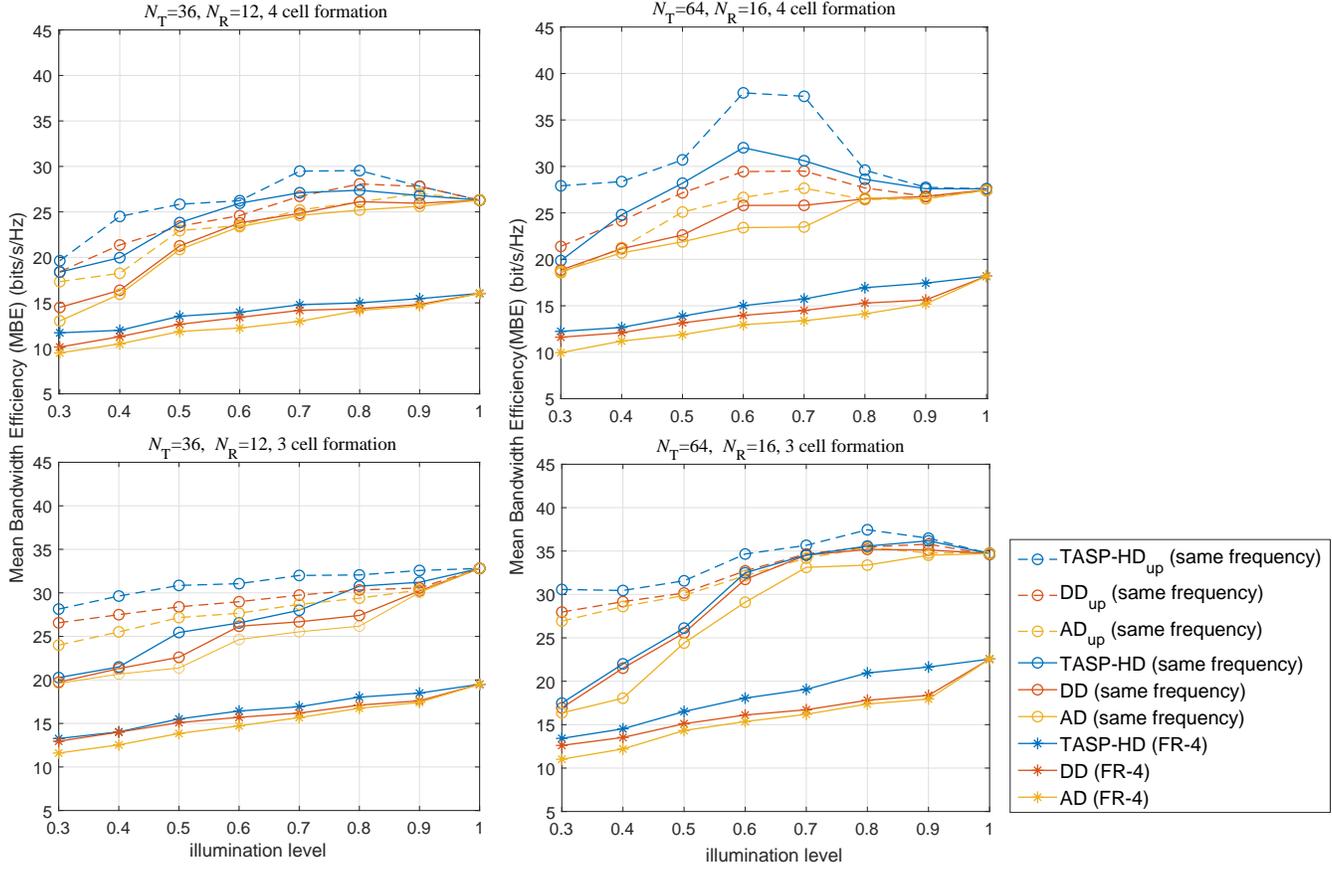}}
	\caption{MBE of TASP-HD and FR under different LED, user and cell formations.}
	\label{figMBE16r64t}
\end{figure*}

The illumination distribution in the indoor space with $N_{\rm{T}}=36,64$ under 60\%, 70\% dimming levels is given in Fig. \ref{fig3Duir}. In order to provide sufficient illuminance, the maximum luminous intensity ${I(0)}$ of each LED is set to 900 and 600 cd when $N_{\rm{T}}=36,64$, respectively. We can observe that in the corners of the indoor space, the illuminance distribution of TASP-HD is more smooth than DD/AD, and the illuminance of TASP-HD is in the range of [300 lx, 1500 lx], which obeys ISO standard\cite{ISOuir}. For example, when $N_{\rm{T}}=64$ and the dimming level is 70\%, the illuminance range of TASP-HD is [311 lx, 1402 lx]. In contrast, the illuminance range of DD/AD is [345 lx, 1916 lx]. In addition, when $N_{\rm{T}}=36$ and the dimming level is 70\%, the illuminance range of TASP-HD is [302 lx, 855 lx] while the illuminance range of DD/AD is [196 lx, 1322 lx]. These results demonstrate that our proposed TASP-HD can achieve better illumination performance.
\vspace{-0.cm}
\subsection{Communication Performance Evaluations}
\begin{figure*}
	\centerline{\includegraphics[width=0.9\linewidth]{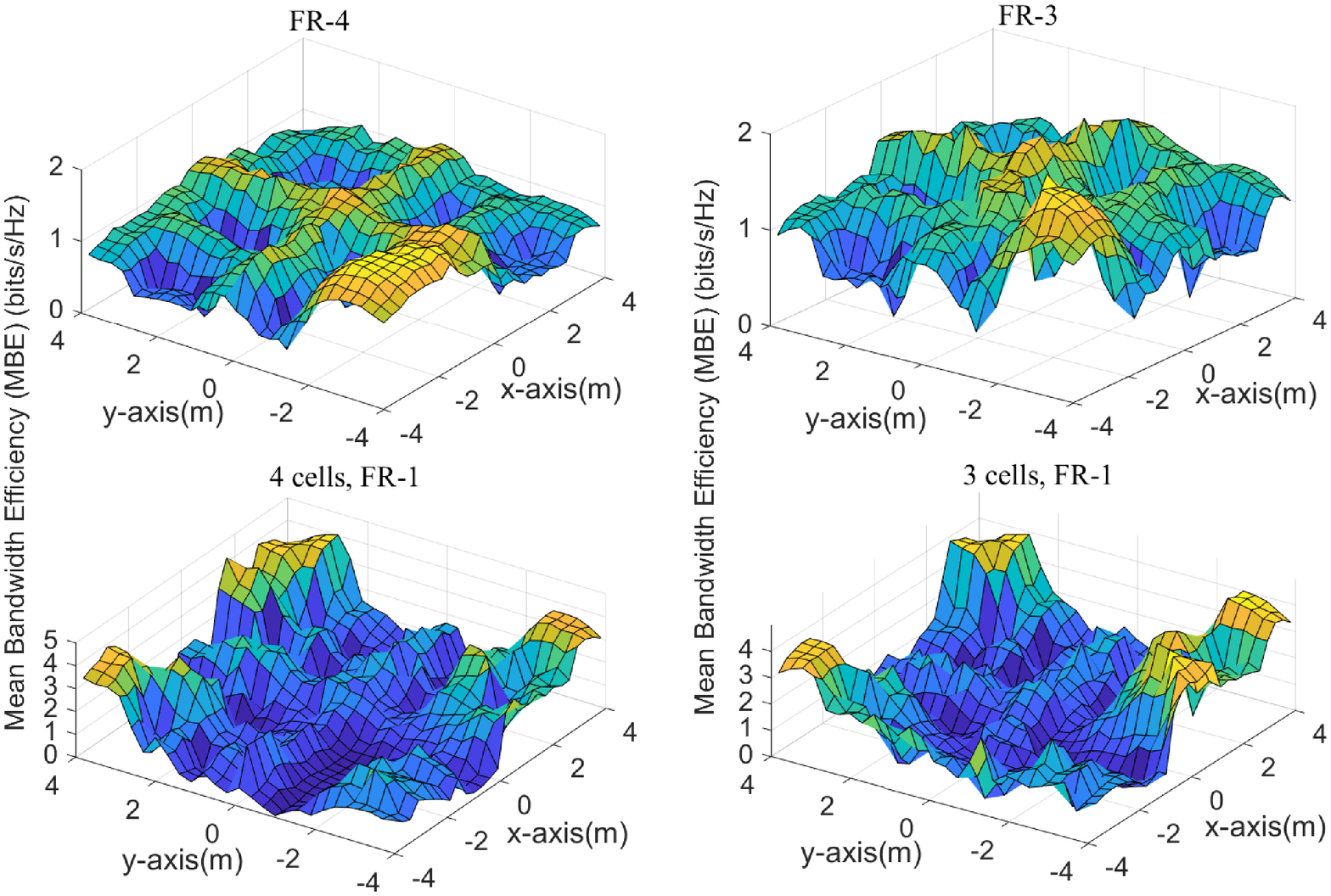}}
	\vspace{-0.cm}
	\caption{MBE distribution of different transmit schemes and cell formations under $\eta=70\%$, $N_{\rm{T}}=64$ and $N_{\rm{R}}=16$.}
	\label{3Dcapa}
\end{figure*}
The communication performance of TASP-HD is evaluated in terms of mean bandwidth efficiency (MBE) \cite{FRCT}. When investigating MBE, frequency reuse (FR) are adopted as baseline scheme. In FR scheme, the bandwidth is divided into $n$ parts, and the LEDs in each cell use different frequencies to transmit signals. In particular, $n=1$ is the special case that the same frequency is used by all cells; $n=N_c$ refers to the case that each of $N_c$ cell has different frequencies, and thus the inter-CI is totally eliminated. In this subsection, we compare FR-1 with FR-3 and FR-4 in 3 cell formation and 4 cell formation, respectively. The MBE of FR with $n \left(n>1\right)$ cell formation is calculated as ${\eta _{{\rm{FR}} - n}} = \frac{R}{n}$, where
$R$ is given by \eqref{sumrateA}.
The MBE performance of TASP-HD, DD, and AD with different LED and user deployments is presented in Fig. \ref{figMBE16r64t}, where TASP-HD, DD, and AD using FR-1, FR-3, and FR-4 are compared. Furthermore, in order to evaluate the trade-off between the sum-rate of all users and the illumination uniformity, we also plot the curves without uniform illumination constraint \eqref{maxsumratec}, and they are termed as TASP-HD$_{\rm{up}}$, DD$_{\rm{up}}$ and AD$_{\rm{up}}$. As shown in Fig. \ref{figMBE16r64t}, the MBE of a system with $N_{\rm{T}}=64$, $N_{\rm{R}}=16$ is higher than a system with  $N_{\rm{T}}=36$, $N_{\rm{R}}=12$. That is because the signal gain increases with the number of LEDs, and thus the SINR is higher with the increased number of LEDs. We can also observe that TASP-HD using FR-1 improves bandwidth efficiency significantly compared TASP-HD with FR-3 and FR-4. For example, when $N_{\rm{T}}=36$, $N_{\rm{R}}=12$ in 4 cell formation under dimming level 80\%, TASP-HD, DD, and AD using FR-1 achieves 12.38 bit/s/Hz, 11.96 bit/s/Hz and 11.05 bit/s/Hz MBE gains compared to that of TASP-HD, DD, and AD using FR-4. Furthermore, TASP-HD always has the best MBE performance among all the considered schemes in different cell formations. In particular, when $N_{\rm{T}}=64$, $N_{\rm{R}}=16$ in 4 cell formation using FR-1, the MBEs of TASP-HD are 4.8 bit/s/Hz and 7.13 bit/s/Hz higher than that of AD, DD under dimming level of 70\%, respectively. This is because the DC bias of AD increases with dimming levels, so the DC bias of AD is lower than DD and TASP-HD. Therefore, according to \eqref{maxsumrateb}, the amplitudes of AD signals are limited, thus reducing the SINR. Besides, TASP-HD can effectively mitigate the side effect of channel correlation existing in DD and AD systems, since it only activates parts of LEDs to transmit signals. As dimming level increases, the number of activated LEDs in TASP-HD approaches $N_{\rm{T}}$, and thus the MBEs of these three schemes approach similar. In particular, when the dimming level is 100\%, TASP-HD, AD and DD achieve the same sum-rate. On the other hand, TASP-HD keeps a balance between illumination uniformity and sum-rate of users, so the MBE of TASP-HD is lower than TASP-HD$_{\rm{up}}$, and the disparity decreases with increasing the dimming level. For example, for $N_{\rm{T}}=36, N_{\rm{R}}=12$ in 3 cell formation, the MBEs of TASP-HD$_{\rm{up}}$ are 7.87 bit/s/Hz, 5.41 bit/s/Hz and 4.07 bit/s/Hz higher than that of TASP-HD when the dimming levels are 30\%, 50\% and 70\%, respectively. When the dimming level is 100\%, TASP-HD$_{\rm{up}}$ and TASP-HD achieve identical MBE. This is because TASP-HD sacrifices the sum-rate to meet the uniform illumination constraint when the number of activated LEDs is small when the dimming level is lower than 50\%, and the illumination uniformity constraint is naturally satisfied with increased number of activated LEDs when the dimming level is higher.

To evaluate the communication performance with different user locations in the indoor space, the MBE performance of one mobile user with TASP-HD using FR-1, FR-3 and FR-4 is shown in Fig. \ref{3Dcapa}. The user moves randomly all over the indoor space while the positions of the rest 15 users are fixed\cite{3dsumrate}. As shown in Fig. \ref{3Dcapa}, TASP-HD with FR-1 significantly improves MBE, especially in the cell center area. In particular, for TASP-HD with FR-1 in 3 cells, the MBEs in the corners of the indoor space (($-4$ m, $-4$ m), ($-4$ m, $4$ m), ($4$ m, $-4$ m), ($4$ m, $4$ m)) are 1.538 bits/s/Hz, 3.26 bits/s/Hz, 3.475 bits/s/Hz and 3.308 bits/s/Hz, respectively, while the MBEs in the same locations for FR-3 are 1.442 bits/s/Hz, 0.96 bits/s/Hz, 1.039 bits/s/Hz and 0.9877 bits/s/Hz, respectively. However TASP-HD with FR-1 suffers from relatively poor performance in cell edges due to the high inter-CI. For example, for TASP-HD with FR-1, the MBE at the center of the indoor space is 1.313 bits/s/Hz, which is greatly lower than cell center areas. Meanwhile, for TASP-HD with FR-3, the MBE at the center location is 1.817 bits/s/Hz, which is close to the MBE in other locations. This is because ZF precoding eliminates intra-CI at the cost of bandwidth efficiency.

\section{Conclusion}
This paper has proposed a TASP-HD scheme for MC MU-MISO VLC systems. The proposed TASP-HD circumvented the challenges of inter and intra-CI by joint design of LED selection and precoding matrix while satisfying illumination constraints, which follows the paradigm of hybrid dimmings. This design problem leads to a non-convex mixed integer optimization problem. To solve this problem, we have divided it into two subproblems, and solved them in an iterative manner. Numerical and simulation results have shown that the proposed TASP-HD improves both the illumination uniformity and sum-rate of users with higher bandwidth efficiency. The mean bandwidth efficiency of TASP-HD is 4.8 bit/s/Hz and 7.13 bit/s/Hz greater than that of AD, DD in a typical indoor scenario under dimming level of 70\%.
\begin{appendices}
\section{Proof of Proposition 2}
We first denote the solution of the $t$th iteration and the optimal value of the second subproblem as $\overline R  = \mathop {\lim }\limits_{t \to \infty } \inf R\left( {{\boldsymbol{q}}_c^{\left( t \right)}} \right)$ and ${R^*} = \inf R\left( {{{\boldsymbol{q}}_c}} \right)$, respectively. To prove the convergence of Algorithm 1, we use proof by contradiction to show that ${R^*} = \overline R$.

We assume a contradiction that $\overline R$ is inferior to ${R^*}$, and thus there exists an $\varepsilon  > 0$ such that
\begin{equation}
{R^*} - 2\varepsilon > \overline R. \label{app2-1}
\end{equation}
Then we have $\widehat {{{{\boldsymbol{q}}_c}}}$ satisfying
\begin{equation}
{R^*} - 2\varepsilon > R\left( \widehat {{{{\boldsymbol{q}}_c}}} \right).  \label{app2-2}
\end{equation}
Let $t_0$ be a large enough iteration, such that for all $t \ge t_0$, we have
\begin{equation}
R\left( {{\boldsymbol{q}}_c^{\left( t \right)}} \right) \ge {R^*} - \varepsilon.\label{app2-3}
\end{equation}
By combining \eqref{app2-2} and \eqref{app2-3}, we have
\begin{equation}
R\left( {{\boldsymbol{q}}_c^{\left( t \right)}} \right)- R\left( \widehat {{{{\boldsymbol{q}}_c}}} \right)>\varepsilon, \forall t \ge t_0. \label{app2-4}	
\end{equation}
Denote the subgradient of $R\left( {{\boldsymbol{q}}_c^{\left( t \right)}} \right)$ as ${\partial R\left( {{\boldsymbol{q}}_c^{\left( t \right)}} \right)}$. Based on Proposition 6.3.1(a) in \cite{bertsekas}, when $t \ge t_0$, we have
\begin{align}
{\left\| {{\boldsymbol{q}}_c^{\left( {t + 1} \right)} - \widehat {{{\boldsymbol{q}}_c}}} \right\|^2} &\le {\left\| {{\boldsymbol{q}}_c^{\left( t \right)} - \widehat {{{\boldsymbol{q}}_c}}} \right\|^2} - 2{\theta ^{\left( t \right)}}\varepsilon  + {\left( {{\theta ^{\left( t \right)}}} \right)^2}{c^2} \notag\\
&= {\left\| {{\boldsymbol{q}}_c^{\left( t \right)} - \widehat {{{\boldsymbol{q}}_c}}} \right\|^2} - {\theta ^{\left( t \right)}}\left( {2\varepsilon  - {\theta ^{\left( t \right)}}{c^2}} \right),\label{app2-5}
\end{align}
where $c$ is a bound of the subgradient, written as $c \ge \sup \left\{ {\left\| g \right\||g \in {\partial R\left( {{\boldsymbol{q}}_c^{\left( t \right)}} \right)}} \right\},\forall t \ge 0$. Since the stepsize of the subgradient algorithm ${\theta ^{\left( t \right)}} = \frac{a}{{\sqrt t }}$ diminishes to zero and satisfies $\sum\limits_{t = 0}^\infty  {{\theta ^{\left( t \right)}} = \infty }$, we have
\begin{equation}
2\varepsilon -	2{\theta ^{\left( t \right)}}{c^2} \ge \varepsilon,\forall t \ge t_0.
\end{equation}
Therefore \eqref{app2-5} can be rewritten as
\begin{align}
{\left\| {{\boldsymbol{q}}_c^{\left( {t + 1} \right)} - \widehat {{{\boldsymbol{q}}_c}}} \right\|^2} &\le {\left\| {{\boldsymbol{q}}_c^{\left( t \right)} - \widehat {{{\boldsymbol{q}}_c}}} \right\|^2} - {\theta ^{\left( t \right)}}\varepsilon \notag\\
& \le  \cdots  \le {\left\| {{\boldsymbol{q}}_c^{\left( {{t_0}} \right)} - \widehat {{{\boldsymbol{q}}_c}}} \right\|^2} - \varepsilon \sum\limits_{j = {t_0}}^t {{\theta ^{\left( j \right).}}},\forall t \ge t_0, \label{app2-6}
\end{align}
which cannot hold, since $\sum\limits_{j = {t_0}}^t {{\theta ^{\left( j \right)}}}  \to \infty$ when $t$ is sufficiently large. Thus the assumption \eqref{app2-1} is a contradiction. Hence we have ${R^*} = \overline R$. The proof is complete.

\section{Proof of Convergence of Algorithm 2}
To prove the convergence of Algorithm 2, we need to prove that the sum-rate of users $R\left( {\boldsymbol{W}}_c,{\boldsymbol{A}}_c \right)$ is nondecreasing in each iteration.

Let $R\left( {{\boldsymbol{W}}_c^{\left( t \right)},{\boldsymbol{A}}_c^{\left( t \right)}} \right)$ denote the sum-rate of users after calculating ${{\boldsymbol{W}}_c^{\left( t \right)},{\boldsymbol{A}}_c^{\left( t \right)}}$ in the $t$th iteration. As we have discussed in Section III, by utilizing an interior-point method to solve the first subproblem \eqref{provequal} we have \cite{penal}
\begin{equation}
R\left( {W_c^{\left( {t - 1} \right)},A_c^{\left( t \right)}} \right) \ge R\left( {W_c^{\left( {t - 1} \right)},A_c^{\left( {t - 1} \right)}} \right).\label{app3-1}
\end{equation}

Meanwhile, for the second subproblem \eqref{sub2withq} based on \textbf{Proposition 2}, we have
\begin{equation}
	R\left( {W_c^{\left( {t} \right)},A_c^{\left( t \right)}} \right) \ge R\left( {W_c^{\left( {t - 1} \right)},A_c^{\left( {t} \right)}} \right).\label{app3-2}
\end{equation}

Moreover, to ensure that the alternate optimization step always improves the objective value, if ${R^{(t + 1)}} < {R^{(t)}}$, we set ${R^{(t + 1)}} = {R^{(t)}}$ as shown in step 7 of Algorithm 2. In that case, the algorithm will terminate since the termination condition ${\left| {{R^{\left( t+1 \right)}} - {R^{\left( t \right)}}} \right|^2} \le {{\varepsilon}_3}$ is satisfied.
Therefore, we can conclude that the value of the objective function must be improved or fixed in each iteration, which is denoted as
\begin{equation}
	R\left( {W_c^{\left( {t} \right)},A_c^{\left( t \right)}} \right) \ge R\left( {W_c^{\left( {t - 1} \right)},A_c^{\left( {t - 1} \right)}} \right). \label{app3-3}
\end{equation}

Given \eqref{app3-3}, Algorithm 2 will finally reach convergence. This completes the proof.
\end{appendices}

\vspace{-0.cm}
\hspace{-10mm}
\bibliographystyle{IEEEbib}
\hspace{-10mm}
\nocite{*}\hspace{-10mm}
\bibliography{TASP-HD_ICI_ref}

\end{document}